\documentclass[%
 reprint,
 amsmath,amssymb,
 aps,
 pre
]{revtex4-2}

\usepackage{graphicx}
\usepackage{dcolumn}
\usepackage{bm}
\usepackage{enumerate}
\usepackage{upgreek}
\usepackage{subfig}
\usepackage{color}
\usepackage{epsf}

\newcommand{\SA}{\mathcal{S}}
\newcommand{\DA}{\mathcal{D}}
\newcommand{\dRM}{\mathrm{d}}
\newcommand{\eps}{{\varepsilon}}
\newcommand{\mv}{{\bm v}}
\newcommand{\mx}{{\bm x}}
\newcommand{\mk}{{\bm k}}
\newcommand{\mq}{{\bm q}}
\newcommand{\mpp}{{\bm p}}
\newcommand{\boldnabla}{{\bm \nabla}}
\def\dRM{\mathrm{d}}
\def\eRM{\mathrm{e}}

\allowdisplaybreaks

\begin{document}

\preprint{APS/123-QED}

\title{Renormalization group study of superfluid phase transition: effect of compressibility}


\author{Michal Dan\v{c}o}
 \email{michal.danco@gmail.com}
\affiliation{%
 Institute of Experimental Physics SAS, Watsonova 47, 040 01 Ko\v{s}ice, Slovakia  
}%

\author{Michal Hnati\v{c}}
 \email{hnatic@saske.sk}
\affiliation{
 Institute of Experimental Physics SAS, Watsonova 47, 040 01 Ko\v{s}ice, Slovakia
}%
\affiliation{
 Bogolyubov Laboratory of Theoretical Physics, Joint Institute
for Nuclear Research, 141980  Dubna, Russian Federation
}%
\affiliation{%
 Faculty of Science, \v{S}af\'arik University, Moyzesova 16, 040 01 Ko\v{s}ice, Slovakia 
}%
\author{Tom\'a\v{s} Lu\v{c}ivjansk\'y}
\email{tomas.lucivjansky@upjs.sk}
\affiliation{%
 Faculty of Science, \v{S}af\'arik University, Moyzesova 16, 040 01 Ko\v{s}ice, Slovakia 
}%
\author{Luk\'a\v{s} Mi\v{z}i\v{s}in}
\email{mizisin@theor.jinr.ru}
\affiliation{
 Institute of Experimental Physics SAS, Watsonova 47, 040 01 Ko\v{s}ice, Slovakia
}%
\affiliation{
 Bogolyubov Laboratory of Theoretical Physics, Joint Institute
for Nuclear Research, 141980  Dubna, Russian Federation
}%

\date{\today}

\begin{abstract}
Dynamic critical behavior 
 in superfluid systems is considered in a presence of
external stirring and advecting processes. 
 The latter are generated by means of the Gaussian random velocity ensemble with white-noise character
 in time variable and self-similar spatial dependence.
  The main focus of this work is to analyze an effect of 
compressible modes on the critical behavior.
 The model is formulated through stochastic Langevin equations, which are then recast into Janssen-De Dominicis response formalism.
  Employing the field-theoretic perturbative renormalization group method we analyze large-scale properties of the model.
Explicit calculations are 
performed to the leading one-loop approximation in the double 
$(\varepsilon,y)$ expansion scheme, where $\varepsilon$ is a deviation
from the upper critical dimension $d_c=4$ and $y$ describes a scaling
properties of the velocity ensemble. 
Altogether five distinct universality classes
are expected to be macroscopically observable. In contrast to the incompressible case, we found that compressibility 
 leads to an enhancement and stabilization of non-trivial asymptotic regimes.
\end{abstract}

\maketitle


\section{\label{sec:intro}Introduction}

Scaling behavior and related concepts arguably provide many fruitful views not only
in theoretical physics~\cite{Tauber2014,HHL08}. Though initially scaling came to prominence in the field of 
high energy physics and critical phenomena, nowadays, many of its applications can be found in such diverse research areas
 as biology~\cite{munoz18,avraham}, finance~\cite{mandelbrot}, population 
dynamics~\cite{krapivsky,Tauber2014}, epidemics spreading~\cite{romualdo15} and others.


Among the most studied systems in physics, exhibiting scaling behavior, is superfluid phase transition in liquid helium. 
In this paper we are concerned with a specific aspect of critical dynamics in the vicinity of $\lambda$-point in superfluid helium  $\mbox{}^4$He.
 In the seminal review of Hohenberg-Halperin~\cite{HH77}, these authors
categorized various dynamical models to which renormalization group methods have been
applied. In order to categorize other forms as well they have denoted
these models in alphabetical order so that E designates a symmetric planar
magnet, while F designates asymmetric planar magnet.
In more general model F, dynamics is captured by three fields. Two of them, $\psi$ and $\psi^\dagger$ correspond to order parameter
and stand for expectation values of microscopic bosonic operators $\hat{\psi}$ and $\hat{\psi}^\dagger$. Third field $m$ describes
 temperature fluctuations in a system. Interactions between fields are determined from the generalized Poisson brackets, whose
 forms follow from physically motivated considerations~\cite{Folk06,Tauber2014}. 
 Model E can be
 interpreted as a simplified version of model F in which certain temperature dependence has been neglected~\cite{Folk06,Vasiliev} and
 physically different variables are empoyed. However, in practical terms this amounts to an appearance
 of a complex kinetic coefficient and an intermode cubic coupling term~\cite{Vasiliev,Folk06}. 
  Both models E and F have been analyzed predominantly by renormalization group (RG) methods~\cite{Folk06,Tauber2014,Vasiliev,mazenko}.
 There remains a long-standing issue~\cite{Folk06,Dohm06,Tauber2014,PRE16,HKMN19b} related to determination, which fixed point
 of the RG flow actually corresponds to a macroscopically observable regime in experiments.
 This is not only an academic problem as ensuing non-asymptotic effects hinder experimentally measurable quantities.
Possible solutions involve (i) search for correct microscopic model for superfluidity~\cite{Komarova,HKMN19b}, (ii) elaborating existing
 numerical results through multi-loop calculations~\cite{Dominicis77,Dominicis78,Dohm79,danco16}, or (iii) appropriate generalizations of models~\cite{PRE16,HKMN19b}.

In this work, we follow the third option (iii) by means of inclusion into the model description external velocity fluctuations.
In this regard several generalization have already been proposed~\cite{krasnov,TMF13,PRE16}.
 It has been shown that incompressible hydrodynamic fluctuations contribute significantly to the value of the $\omega_w$-index,
 which controls stability of large-scale regimes~\cite{Folk06,Dominicis78}. This index is related to a RG behavior
 of a ratio of two kinetic coefficients (in this paper corresponding ratio is related to a parameter $u$ introduced later in Sec.~\ref{sec:problem}).  
  
 However, the overall conclusions are by no means
 decisive. Main problem we want to address in this paper is to analyze a presence of 
 solenoidal modes in velocity fluctuations. In particular, we study what new effects in comparison with the incompressible case can be expected
  and to what extent is critical behavior affected. 

Similarly to typical advective problems in fluid dynamics~\cite{Frisch,turbo,FGV01} we incorporate
 velocity fluctuation field $\mv$ by substituting partial time derivative
$\partial_t$ with a convective derivative of the form $\partial_t+({\mv}\cdot\boldnabla)$.
 From general considerations~\cite{ivanov}, we expect that presence of external disturbance, e.g. random impurities or turbulent mixing,
  might lead to completely new types of critical behavior with richer and more
exotic properties~\cite{KH, JOS, SR, SR1}. 
 
To fully specify a theoretical setup let us briefly describe the employed model for velocity fluctuations. 
 We assume that velocity  $\mv=\mv(t,\mx)$ is a random stochastic Gaussian variable with prescribed statistical properties~\cite{FGV01,Ant06}.
In the original formulation~\cite{Kraichnan68}, the velocity field was further taken to be isotropic, incompressible 
and decorrelated in a time variable. 
Without loss of generality, we can set the mean value $\langle \mv \rangle = 0$ and take 
 the pair velocity function in the following form 
\begin{equation}
  \langle v_iv_j \rangle \propto \delta(t-t') k^{-d - y} T_{ij},
\end{equation}  
   where $k$ is the wave number, $0<y<2$ is a free parameter with 
the realistic (“Kolmogorov”) value 
 $y= 4 / 3$, $d$ is a space dimension, and tensor $T_{ij}$ carries information about
 vectorial character of velocity modes.
 This model attracted a lot of interest in the past mainly because 
of insights it offers
into the origin of intermittency and anomalous scaling in the fully developed turbulence~\cite{FGV01,Ant06}. 
Naively, basic premises of such models might be perceived as too crude and unrealistic. Nevertheless, important effects of 
parity breaking, anisotropy, or compressibility are easily taken into account~\cite{Ant99, AAH02, Ant00,AHHJ03,Ant06}. It turns out that then the phenomenon
of intermittency is even more pronounced than in genuine turbulent flow.
 The recent studies have also pointed out some significant differences between the zero and finite correlation 
time problems \cite{falko, eyink, Ant00} and between the compressible and incompressible cases \cite{verga, AdzAnt98}. 

Let us point out a crucial difference between critical dynamical models and the model considered in this work.
The basic assumption of the former models is a presence of ambient thermal fluctuations. Coupling with
thermal bath provides necessary means by which a critical steady state can be maintained. Deviations from thermal
equilibrium are considered small that results into variety of relations between different physical quantities~\cite{mazenko}.
 A well-known example is the fluctuation-dissipation theorem, which relates two-point correlation function to susceptibility~\cite{Tauber2014,Vasiliev}.
 On the other hand, inclusion of external velocity fluctuations effectively drives the critical system away
from the thermal equilibrium and leads to an effectively non-equilibrium system. Hence, relations like the fluctuation-dissipation theorem cease to hold, and
 as a consequence, a theoretical analysis becomes more involved.

In relation to this paper, there was recently put forward an intriguing approach to a similar problem.
In contrast to the standard approach to critical systems, in which dynamical models are constructed using generalized Poisson brackets or
symmetry considerations~\cite{Folk06}, particular microscopic approach was suggested~\cite{honkonen13,HKMN19b,HKMN19c}.
The authors have analyzed various  aspects of phase transitions in superfluids were analyzed by means of
 a non-trivial technique of non-equilibrium Green functions~\cite{Vasiljev1}.
 In particular, 
 an implicit assumption common to many critical models related to incompressibility of
underlying fluid~\cite{Landau_fluid} was put in a question. Relaxing this condition with allowance of compressible modes results into an effective model
fully equivalent to model A of critical dynamics~\cite{HH77}. This result seems peculiar as
model A is conceivably the simplest dynamical extension of well-known $\varphi^4$ model~\cite{Zinn,Vasiliev}. For
 one-component order parameter takes simple form
\begin{equation}
	 \SA_\text{A} = -\int\!\! \mathrm{d}^d x\! 
	  \left(
	 \frac{1}{2}(\boldnabla\varphi)^2  + \frac{\tau}{2} \varphi^2 +  
	\frac{g}{4!} \varphi^4 \right),
	\label{eq:modelA}
\end{equation}
 also known in the literature as Landau-Ginzburg-Wilson action functional~\cite{Zinn,Vasiliev,Tauber2014}.
As experiments are still lacking in this direction, suggested models are still waiting for a decisive affirmation of their relevance for
critical dynamics.
 
The paper is organized as follows. In Sec.~\ref{sec:problem} we give a formulation of the problem by means of Langevin equations.
These are then rewritten into the field-theoretic model using Janssen-De Dominicis formalism.
 The resulting action is amenable to the field-theoretic  renormalization group analysis, which 
 is carried out in  Sec.~\ref{sc:renormalization}.
  Sec.~\ref{sec:regimes} is devoted to a  detailed analysis of fixed points' structure, and Sec.~\ref{sec:crit_dim} is reserved
for a calculation of experimentally relevant critical exponents. Concluding remarks are summarized in Sec.~\ref{sec:conclusion}. 
Supplementary sections \ref{sec:app_feynman}, \ref{sec:app_anomdim} and \ref{app:coord} contain technical details about 
divergent parts of Feynman diagrams, lengthy expressions of RG functions and coordinates of fixed points.

 {\section{Field-theoretical formulation} \label{sec:problem}}
Using the standard terminology proposed in \cite{HH77}, model E of critical dynamics is
described by the non-conserved two-component order parameter composed of two (complex)
conjugated fields $\psi(t,{\bm x})$ and $ \psi^{\dagger}(t,{\bm x})$, and a conserved scalar
field $m(t,{\bm x})$. The former can be viewed as macroscopic averages of the Bose-particle field operators, whereas the 
latter field $m(t,{\bm x})$ is a certain linear combination of energy and mass density \cite{Vasiliev} (or a normal
component of the magnetization in antiferromagnetic materials). Time evolution of the fields is governed
\cite{Vasiliev,Folk06,Tauber2014} by the following set of equations
\begin{align}
      \partial_t \psi & =   \lambda_0\frac{\delta \SA_{st}}{\delta \psi^{\dagger}} + 
      i\lambda_0 g_{30}\psi\frac{\delta \SA_{st}}{\delta m} + f_{\psi}, 
      \label{eq:S1}
       \\
      \partial_t \psi^{\dagger} & =  \lambda_0\frac{\delta \SA_{st}}{\delta \psi} -
      i\lambda_0 g_{30}\psi^{\dagger}\frac{\delta \SA_{st}}{\delta m}
       + f_{\psi^{\dagger}}, 
       \label{eq:S2}
       \\
      \partial_t m & =   - \lambda_0 u_0 \boldnabla^2 \left(\frac{\delta \SA_{st}}{\delta m}\right) + 
      i\lambda_0 g_{30}\left(\psi^{\dagger}\frac{\delta \SA_{st}}{\delta \psi^{\dagger}} - 
      \psi \frac{\delta \SA_\text{st}}{\delta \psi}\right) \nonumber\\
      & + f_m,
      \label{eq:S3}
\end{align}
where $\boldnabla^2 = \sum_{i=1}^d \partial^2/\partial x_i\partial x_i$
is Laplace operator in $d$-dimensional space, $\partial_t=\partial/\partial t$ is time derivative, $\lambda_0$ is a 
kinetic coefficient related to diffusive modes. Let us note that when necessary, we write space dimension $d$ explicitly.
This is due to a later use of the RG approach. 
In contrast to action~\eqref{eq:modelA}
 now the static action functional $\SA_\text{st}$ is given by the following form 
in the critical region
\begin{equation}
	 \SA_\text{st} = \int\!\! \mathrm{d}^d x\! \left(\psi^{\dagger}
	\boldnabla^2 \psi - \frac{1}{2}m^2 + m h_0 - 
	\frac{1}{6}g_{10}(\psi^{\dagger}\psi)^2 \right),
	\label{eq:static}
\end{equation}
and substitution $\varphi(\mx)\rightarrow\varphi(t,\mx)$ is implicitly assumed in Eq.~(\ref{eq:S1})-(\ref{eq:S3}) in 
terms stemming from variational derivatives for any member from the set $\varphi \in \{ \psi,\psi^\dagger,m \}$. 
Parameters $g_{10}$ and $g_{30}$ 
play a role of
coupling constants of the theory \cite{Vasiliev,Tauber2014}. Random forces $f_{\psi}$, $f_{\psi^\dagger}$ and $ f_m$ are
assumed to be Gaussian random variables
with zero means and correlators $D_{\psi}$, $D_{\psi^\dagger}$ and $ D_m$ with the white noise character in a time
variable. In the time-momentum representation they are given by the following formulas
\begin{align}
	D_{\psi^\dagger}(p,t,t') & = D_{\psi}(p,t,t') = \lambda_0 \delta(t-t'), 
	\label{eq:correlators1}
	\\
	D_m(p,t,t') & = \lambda_0 u_0 p^2 \delta(t-t').
	\label{eq:correlators2}
\end{align}
Parameter $u_0$ is dimensionless and has been  introduced for future convenience.
Here and below, the bare (unrenormalized) parameters in the renormalization group sense are denoted with the
subscript ''0''. The normalization in relations (\ref{eq:correlators1})-(\ref{eq:correlators2}) has been chosen 
in such a way that the steady-state equal-time correlation functions of the stochastic problem are calculable
exactly with the Boltzmann factor
 $\exp(\mathcal{S}_\text{st})$. 
 The stochastic problem (\ref{eq:S1}-\ref{eq:S3}), 
(\ref{eq:correlators1}), and (\ref{eq:correlators2}) can be concisely reformulated by means of
 De Dominicis-Janssen functional formalism \cite{Jan76,Dom76}. Ensuing field-theoretic action of model E
 \cite{Vasiliev,Tauber2014,Folk06} then directly follows
\begin{align}
	 \SA_\text{E} & =  2\lambda_0 {\psi ^{\dagger}}'\psi
	'-\lambda_0 u_0 m' \boldnabla^2 m' 
	 + {\psi^{\dagger}}'\{-\partial_t\psi + 
	\lambda_0 [\boldnabla^2 \psi \nonumber \\ 
	& - g_{10}(\psi^{\dagger}\psi)\psi/3 ] +i\lambda_0
	g_{30} \psi[-m+h] \} +  \mbox{H. c.} \nonumber\\ 
	& +  m'\{-\partial _tm -\lambda_0 u_0 \boldnabla^2 [-m+h]
	\nonumber\\
	& + i\lambda_0 g_{30}[\psi^{\dagger} \boldnabla^2 \psi-\psi {\boldnabla^2}\psi^{\dagger}]\}.
	\label{eq:action}
\end{align}
Abbreviation H.~c.~stands for a Hermitian conjugate part of the action with respect to the $\psi$-field. In 
action \eqref{eq:action} we have employed a condensed notation, in which
integrals over space-time are implicitly included. For instance, the second 
term in (\ref{eq:action}) is an abbreviated form of the expression
 $  m'\partial ^2m' = \int\!\! \dRM t\int\!\! \mathrm{d}^d x \mbox{ }m'(t,\mx) \boldnabla^2 m'(t,\mx). $
 Prime fields $\psi'$ and ${\psi^\dagger}'$ correspond to auxiliary Martin-Siggia-Rose response fields \cite{MSR}.
 A functional formulation effectively means that the statistical averages of the random quantities in the original stochastic
 problem (\ref{eq:S1}-\ref{eq:S3}) can be represented by functional integrals over the full set of fields with the weight 
 functional 
  $\exp({\SA}_\text{E} )$.
   In quantum-field-theory terminology  various correlation functions then correspond
 to Green functions of the field theoretic model with action (\ref{eq:action}).
 Such formulation is especially convenient for the further use of field-theoretical methods such as
 Feynman diagrammatic technique and
 perturbative
 renormalization group, which provide main theoretical tools in this work.

The next step consists in an introduction of the velocity fluctuations into a theoretical model. According to a
standard approach \cite{Landau_fluid,turbo,Ant06} it is sufficient to replace the partial time derivative 
$\partial_t$ by the Lagrangian derivative 
$\partial_t + (\mv \cdot \boldnabla)$. However, in  presence of compressibility 
this is not sufficient \cite{AK10}, and the following substitutions are necessary
\begin{align}
  \partial_t \psi &\rightarrow \partial_t \psi + (\mv \cdot \boldnabla) \psi + a_{10} \psi(\boldnabla \cdot \mv ),
  \label{eq:replace1} 
  \\
  \partial_t m &\rightarrow \partial_t m + (\mv \cdot \boldnabla) m + a_{20} m(\boldnabla \cdot \mv). 
  \label{eq:replace2}
\end{align}
Without inclusion of terms proportional to parameters $a_{10}$ and $a_{20}$ the model ceases to be multiplicatively
renormalizable.

In this work we employ the Kraichnan rapid-change model \cite{Kraichnan68,AdzAnt98,Ant06} with compressibility of the fluid 
taken into account. Accordingly, the velocity field $\mv$ is assumed to be a random Gaussian variable with prescribed
statistical properties. By a proper substitution we can always achieve that $\langle\mv\rangle = 0$. Due to the Gaussian
character of $\mv$ the only needed information lies in a specification of the two-point correlation function, which 
assumes the following form
\begin{equation}
	\langle v_i(t,\mx) v_j(t',\mx^\prime) \rangle = \delta(t-t')D_{ij}( \mx - \mx^\prime ), 
	\label{eq:correlator}
\end{equation}
where Dirac delta function ensures the Galilean invariance of model \cite{turbo}. Due to the translational invariance 
of the flow it is convenient \cite{Ant06} to specify the kernel function $D_{ij}$ in Eq.~\eqref{eq:correlator} in the
Fourier representation
\begin{equation}
	D_{ij}({\bm r}) = D_0\! \int\! \frac{\dRM^d k}{(2\pi)^d} 
	\frac{\theta(k-l_v)}{k^{d+y}}[P_{ij}(\mk) + \alpha Q_{ij}(\mk)] \eRM^{ i \mk\cdot{\bm r} },
	\label{eq:dij}
\end{equation}
where $P_{ij}(\mk) = \delta_{ij} - k_i k_j/k^2$ and $Q_{ij}(\mk) = k_i k_j/k^2$ are the transverse and the longitudinal 
projector, respectively. Further, $k = |\mk|$ is the wave number, $D_0 >0$ is an amplitude factor and $\alpha \ge 0$ is
an arbitrary parameter, which might be interpreted as a degree of compressibility in the system \cite{Ant00,gaw00}.
Heaviside function $\theta(x)$ ensures infrared (IR) cutoff of the theory that does not violate Galilean invariance. 
 Momentum IR scale $l_v$ is related to the external scale of velocity fluctuations $L$ crudely as $l_v \sim 1/L$, but precise
 form is unimportant for the later discussion.

 The case $\alpha=0$ 
corresponds to the incompressible fluid $({\boldnabla} \cdot \mv = 0)$, whereas  $\alpha> 0$ describes a deviation from the
incompressibility.  After a proper rescaling, the limit 
$\alpha \rightarrow \infty$ at  fixed $\alpha D_0$ yields purely potential velocity field. The exponent $0<y<2$ is a free
parameter that might be interpreted as the H\"older exponent, which expresses a roughness of the velocity field. The
Kolmogorov regime corresponds to the value $y = 4/3$, whereas the Batchelor limit (smooth velocity) is obtained in the limit
$y \rightarrow 2$. 

The action functional describing statistics of the velocity field $\mv$ is simply given by a quadratic form
\begin{equation}
  {\SA}_\text{vel}=-\frac{1}{2} v_i D_{ij}^{-1}v_j, 
  \label{eq:action_vel}
\end{equation}
where $ D_{ij}^{-1} $is the kernel of the inverse linear operator in Eq.~(\ref{eq:dij}).
 This yields a propagator $\Delta_{vv}$, which in the time - momentum representation
 takes the following form
\begin{equation}
  \Delta_{vv} (t,\mk) = w_0\lambda_0 \delta(t) \frac{P_{ij}(\mk) + \alpha Q_{ij}(\mk)}{k^{d+y}}.   
  \label{eq:propa6}
\end{equation}
For convenience, the factor $D_0$ from the kernel~(\ref{eq:dij}) has been expressed in the following way 
$D_0 = w_0 \lambda_0 $, so that that  RG constants might depend only on $w_0$.

To summarize, the total dynamic functional for model E with an inclusion of external velocity fluctuations is given by a sum 
of expressions (\ref{eq:action}) and (\ref{eq:action_vel}), i.e.,
\begin{equation}
   {\SA} = {\SA}_\text{E} + {\SA}_\text{vel}.  
   \label{eq:total_act}
\end{equation}
 
Model (\ref{eq:total_act}) is amenable to the standard Feynman diagrammatic technique, which is based on the graphical
interpretation of linear (solvable) part of the action and non-linear terms therein \cite{Tauber2014,Vasiliev}. In graphical
means interaction terms are represented by  vertices, which are connected by lines. The latter correspond to propagators of 
the free theory, which are given by the quadratic part of the action. Propagators are conveniently given in the 
frequency-momentum representation 
\begin{align}
  \Delta_{mm} & = \frac{2\lambda_0 u_0 k^2}{\omega^2 + \lambda_0^2 u_0^2 k^4} \theta(k-l_m), 
  \label{eq:propa1} \\
  \Delta_{mm'} &  = \frac{1}{-i\omega + \lambda_0 u_0 k^2}, 
  \label{eq:propa2}   \\
  \Delta_{\psi' \psi^{\dagger}} &= \Delta_{\psi^{{\dagger}'} \psi} = \frac{1}{i\omega + \lambda_0 k^2}, 
  \label{eq:propa3} \\
  \Delta_{\psi \psi^{{\dagger}'}} & = \Delta_{\psi^{\dagger} \psi'} = \frac{1}{-i\omega + \lambda_0 k^2}, 
  \label{eq:propa4}
  \\
  \Delta_{\psi \psi^{\dagger}} &= \Delta_{\psi^{\dagger} \psi} = \frac{2\lambda_0}{\omega^2 + \lambda_0^2 k^4}\theta(k-l_\psi),
  \label{eq:propa5}
\end{align}
where $l_m$ and $l_\psi$ are IR cutoff scales for fields $m$ and $\psi$. For practical reasons we assume
\begin{equation}
  l_v=l_m=l_\psi \equiv l
  \label{eq:IRscale}
\end{equation}
 in actual evaluations of Feynman diagrams.
 This choice can be adopted as univesal quantities do not depend on a particular choice of IR regularization~\cite{Zinn,Vasiliev}.

With every interaction vertex, the algebraic factor 
 \begin{equation*}
  V_N(x_1,\ldots,x_N;\varphi) = 
  \frac{\delta^N \SA [\varphi]}{\delta\varphi(x_1)\ldots\delta\varphi(x_N)}    
  \label{eq:ver_factor}
\end{equation*}
is associated \cite{Vasiliev}, and $\varphi$ is any field of the theory, i.e. $\varphi \in\Phi $, where 
\begin{equation}
  \Phi = \{\psi, \psi', \psi^{\dagger}, \psi^{\dagger'}, m, m', v \}.
  \label{eq:all_fields}
\end{equation}
Here, we readily find three vertex factors $V_{\psi^{{\dagger}'}\psi^{\dagger} \psi \psi}$, $V_{\psi^{{\dagger}'} \psi m}$, 
$V_{m' \psi^{\dagger} \psi}$ plus their complex conjugates. Their explicit form can be easily inferred from 
action (\ref{eq:action}) and in the frequency-momentum representation, it explicitly reads 
\begin{align}
   V_{\psi^{{\dagger}'}\psi^{\dagger} \psi \psi} & = -\frac{2g_{10}\lambda_0}{3},
   \label{eq:ver_factor1}\\
   V_{\psi^{{\dagger}'} \psi m} & = -\lambda_0 g_{30},
   \label{eq:ver_factor2}
   \\   
   V_{m' \psi^{\dagger}(\mk) \psi(\mq)} & = i\lambda_{0}g_{30}[\mk^2-\mq^2].
   \label{eq:ver_factor3}
\end{align}
The last vertex factor displays a nontrivial dependence on inflowing momenta of fields $\psi^\dagger$ and $\psi$.

In addition, set of propagators (\ref{eq:propa1})-(\ref{eq:propa3}) has to be supplemented with the velocity propagator
$\Delta_{vv}$ defined through the relations (\ref{eq:correlator}) and (\ref{eq:dij}), respectively. Novel interaction vertices arise from the
convective terms~(\ref{eq:replace1})-(\ref{eq:replace2}) as well. Their vertex factors are
\begin{align}
  V_{\psi^{{\dagger}'} \psi(\mk) v_i(\mq)} & = V_{\psi^{'} \psi^{\dagger}(\mk) v_i(\mq)} = ik_i + ia_{10}q_i, 
   \label{eq:ver_factor4}  \\
  V_{m' m(\mk)  v_i(\mq)} & = ik_i + ia_{20}q_i.    
   \label{eq:ver_factor5}
\end{align} 
Let us recall that the parameter $y$ is not related to the spatial dimension and
can be varied independently. For the RG analysis of the full-scale problem it is important that all the interactions become 
logarithmic simultaneously. Otherwise, one of them would be IR irrelevant with respect to the other and it should be
discarded \cite{Vasiliev,Zinn}. 
As a result, some of the scaling regimes of the full model would be lost. 
 Instead of the ordinary $\varepsilon$ expansion
in the single-charge models, the coordinates of the fixed points, critical dimensions and other quantities are now calculable
in double expansion scheme $(\varepsilon, y)$.

\begin{figure}
   \includegraphics[width=8.5cm]{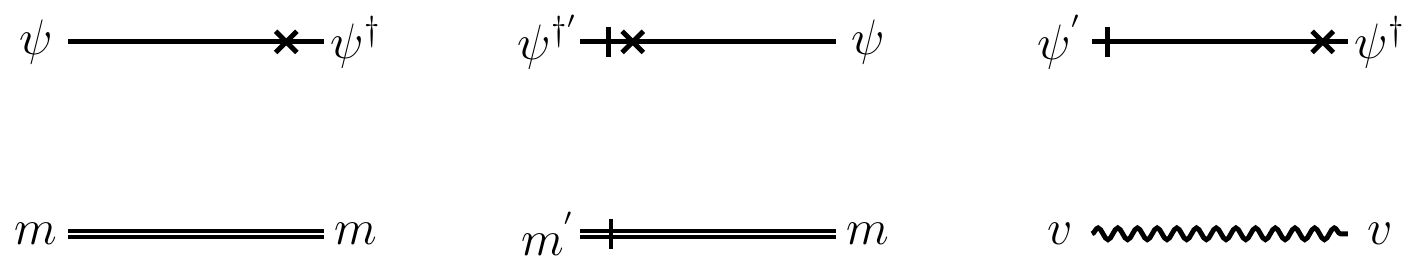}
   \caption{A graphical representation of the free part of the action~\eqref{eq:total_act} that corresponds to lines in 
   the Feynman diagrammatic technique.}
   \label{fig:props}
\end{figure}

\begin{figure}
   \includegraphics[width=8.5cm]{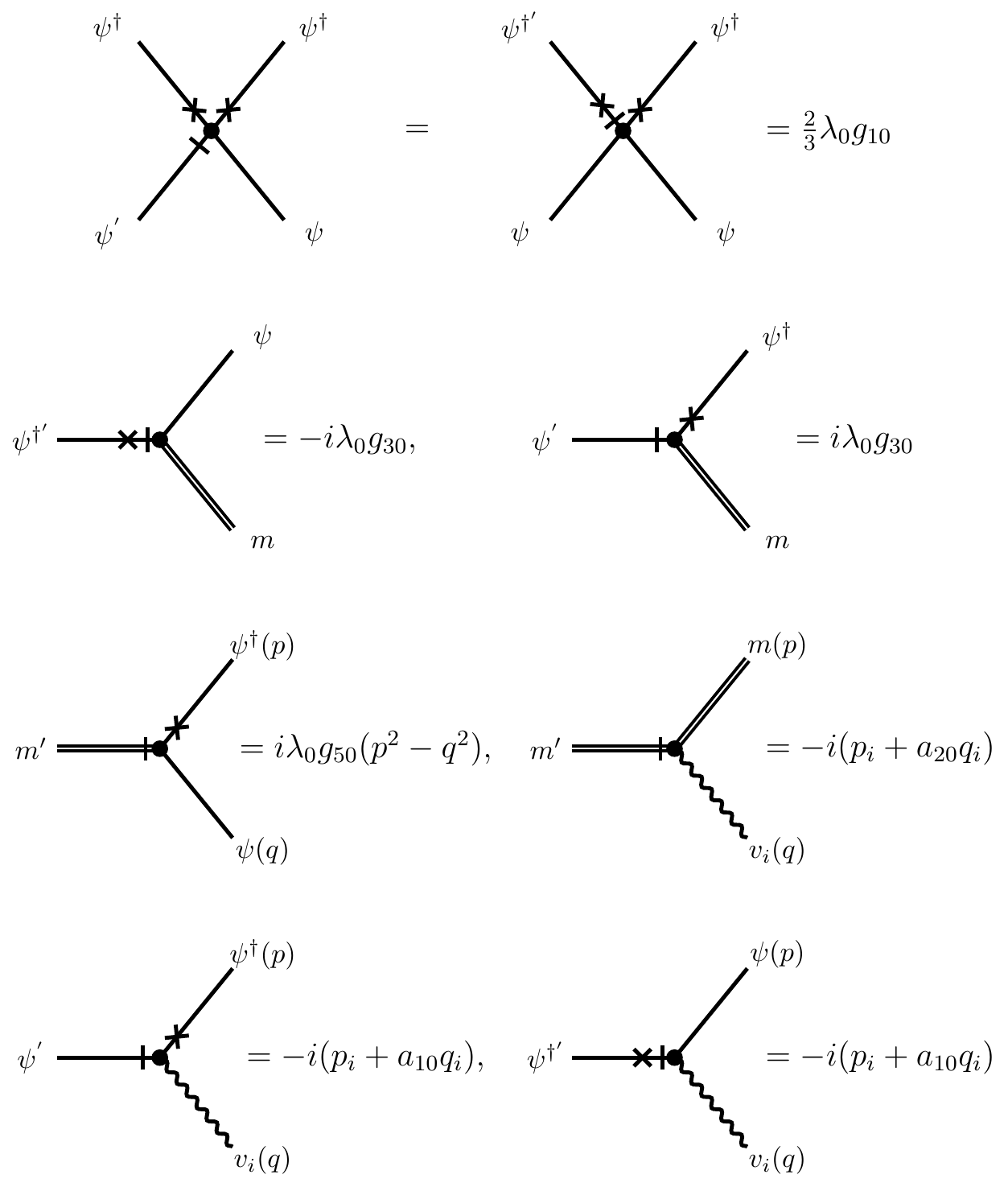}
   \caption{A graphical representation of the nonlinear part of the action~\eqref{eq:total_act} that corresponds to 
   interaction vertices in the Feynman diagrammatic technique.}
   \label{fig:vertex}
\end{figure}

The perturbation theory of the model is amenable to the standard Feynman diagrammatic expansion \cite{Amit,Zinn,Vasiliev}.
A starting 
point of the perturbation theory is a free part of the action (\ref{eq:total_act}). By graphical means, it is represented as 
lines in the
Feynman diagrams, whereas the non-linear terms  in (\ref{eq:total_act}) correspond to vertices
connected by the  lines. 
 The bare propagators are graphically depicted in Fig.~\ref{fig:props}, and interaction vertices in 
Fig.~\ref{fig:vertex}.

{\section{Renormalization group analysis} \label{sc:renormalization}}

A standard goal in statistical physics lies in determination of macroscopic (large-scale) behavior of the system. The RG
procedure allows 
to exploit scale invariance at the critical point and an elimination of UV divergences yields an information about the
IR behavior \cite{Vasiliev,Zinn}. There
are different prescriptions for the renormalization procedure. 

In contrast to the usual situations in critical models, here we deal with a model 
exhibiting two small parameters $\varepsilon$ and $y$. Similarly, it occured in various context in the past~\cite{weinrib83,satten85,VN96}.
Due to presence of two formally small expansion parameters, the RG approach differs somewhat from an usual formalism.
First, we assume that the model is regularized by means of an analytic regularization augmented with a dimensional regularization.
As has been elucidated~\cite{AHK10,honkonen_chaos} mostly used MS scheme suffers from potential deficiencies and is thus not
 satisfactory from a theoretical point of view. Instead,  for calculations 
 of RG constants, we choose a normalization point scheme.
 Because we restrict ourselves here to the leading one-loop approximation, only two types of UV singularities arise: we find either a pole of type $1/\eps$ or
  $1/y$, respectively. Such simple structure pertains 
only to the lowest orders in a perturbation scheme. In higher loop-approximations, poles in the form of general linear combinations 
in $\eps$ and $y$ are 
expected to arise. Moreover, non-trivial issues related to vector character of fields are expected~\cite{wiese97}.

\subsection{Canonical dimensions}

A starting point of the RG approach is an analysis of canonical dimensions. Dynamical models 
of type (\ref{eq:action}), in contrast to static models, demonstrate two-scale behavior. This accounts for an assignment 
of two independent 
(momentum and frequency) canonical dimensions to each quantity $F$ (a field or a parameter in the action functional).
Further, since we work with the translationally 
invariant theory it is sufficient to analyze only one-particle-irreducible (1PI) functions of the model.
 
 The
momentum dimension $d_F^k$ and the 
frequency dimension $d_F^{\omega}$ are determined from the standard normalization conditions 
\begin{align}
  d_k^k & = - d_x^k = 1,  &d_{\omega}^k& = d_t^k = 0, \nonumber \\
  d_k^{\omega} & = d_x^{\omega} = 0,  &d_{\omega}^{\omega}& = - d_t^{\omega} = 1,
  \label{eq:dimens}
\end{align}
and from the requirement that each term in the action functional has to be a dimensionless
quantity \cite{Vasiliev,Zinn}. Then, based on $d_F^k$ and $d_F^{\omega}$ one can introduce a total canonical
dimension $d_F = d_F^k + 2d_F^{\omega}$, (in the 
free theory time derivative $\partial_t$ should scale in the same way as Laplace operator $\partial^2$).
\begin{table}[h]
\small\addtolength{\tabcolsep}{0pt}
	
		\begin{tabular}{|c|c|c|c|c|c|c|}
			\hline
			&&&&&&\\[-1.5mm]
			$Q$ & $p$, $1/x$ & $\omega$, $1/t$ & $\psi, \psi^{\dagger}$ & $\psi',\psi^{{\dagger}'}$ & $m, m',h$ & $v$  
			\\[0.5mm] \hline \hline 
			&&&&&&\\[-1.5mm]
			$d_Q^p$ &1&0& $\frac{d}{2} - 1$ & $\frac{d}{2} + 1$ & $\frac{d}{2}$ & $-1$   
			\\ [0.5mm]\hline 
			&&&&&&\\[-1.5mm]
			$d_Q^\omega$&0 &1& $0$ & $0$ & $0$ & $1$  
			\\[0.5mm]  \hline
			&&&&&&\\[-1.5mm]
			$d_Q$ & 1&2&$\frac{d}{2} - 1$ & $\frac{d}{2} + 1$ & $\frac{d}{2}$ & $1$  
			\\[0.5mm]  \hline   \hline
			&&&&&&\\[-2.mm]
			$Q$ &$\lambda_0$ & $u_0$ & $g_{10}$ & $g_{30}, g_{50}$ & $w_0$&$a_{10}, a_{20}, \alpha$ 
			\\[0.5mm] \hline \hline 
			&&&&&&\\[-1.5mm]
			$d_Q^p$& $-2$ & $0$ & $\eps$ & $\frac{\eps}{2}$ & $ y $&$0$
			\\ [0.5mm]\hline 
			&&&&&&\\[-1.5mm]
			$d_Q^\omega$&$1$& $0$ & $0$ & $0$ & $0$&$0$
			\\[0.5mm]  \hline
			&&&&&&\\[-1.5mm]
			$d_F$ &$0$& $0$ & $\eps$ & $\frac{\eps}{2}$ & $ y $ &$0$
			\\[0.5mm]  \hline
		\end{tabular}
		\caption{Canonical dimensions of the fields and parameters for
			model E~with activated  hydrodynamic modes defined by action~\eqref{eq:total_act}.}
		\label{tab:can_dim}
	
\end{table}
The dimensions of all quantities appearing in action functional ${\SA}$ are summarized in Tab.~\ref{tab:can_dim}. It follows that 
the model is logarithmic (the coupling
constants become dimensionless) when $4-d=0 (\varepsilon = 0)$ and $y=0$. The total canonical dimension of an arbitrary 
1PI function \cite{Vasiliev} is given
by the relation
\begin{equation} 
 d_{\Gamma} = d + 2 -  \sum_{\varphi\in\Phi} N_{\varphi} d_{\varphi},
 \label{eq:total_canon}
\end{equation} 
where the sum runs over a set of all fields $\Phi$ (defined in~\eqref{eq:all_fields})
appearing in a given 1PI function $\Gamma$. The total dimension $d_{\Gamma}$ is a formal index of the UV divergence. Due to 
the compressibility, the derivative
$\partial$ on the external line $m'$ in graphs of the 1-irreducible functions can not be singled out. It follows that the 
formal and real UV exponent are the
same $d_{\Gamma} = \delta_{\Gamma}$. Superficial UV divergences, whose removal requires counterterms, can be present only in
those functions $\Gamma$ for
which $d_{\Gamma}$ is a non-negative integer \cite{Vasiliev}. It is easy to verify that all needed counterterms have a form
of various terms already contained 
in action (\ref{eq:action}). By inspection of the graphs we observe that all nontrivial diagrams in
the term $m' \partial_t m$  vanish. In  \cite{krasnov} it has been demonstrated that the linkage to critical statics is
violated due to the inclusion of the 
velocity field, but the multiplicative renormalization can be recovered by considering a new charge associated
with the interaction term $m'(\psi^{\dagger} \boldnabla^2 \psi - \psi \boldnabla^2 \psi^{\dagger})$. More precisely, instead
of writing this term with the charge
$g_3$ (see Eq. (5.149) in Chapter 5.20 in \cite{Vasiliev}), it has to be given as follows
\begin{equation}
   g_5 m'(\psi^{\dagger} \boldnabla^2 \psi - \psi \boldnabla^2 \psi^{\dagger}),
\end{equation}
where in general a new charge $g_5$ does not coincide with the charge $g_3$, i.e. $g_5\neq g_3$. 
In summary, the field-theoretic renormalized action for model E with velocity fluctuations in
condensed notation takes the following form
\begin{align}
  {\SA}_\text{R} & =  2 Z_1 \lambda {\psi ^{\dagger}}'\psi
     '-Z_2\lambda u m'\partial ^2m'  -\frac{1}{2} v D^{-1}v \nonumber \\ 
     & + {\psi^{\dagger}}'\{-Z_3\nabla_t\psi - Z_4 a_1( \boldnabla \cdot \mv )\psi + 
     \lambda [Z_5 \partial^2\psi \nonumber\\
     &-Z_6 g_1\mu^{\varepsilon}(\psi^{\dagger}\psi)\psi/3 ] 
      + Z_7i\lambda
     g_3\mu^{\varepsilon/2} \psi[-m+h] \}  + \mbox{H. c.}
     \nonumber\\
     & +  m'\{-Z_8\nabla _t m - Z_9a_2( \boldnabla \cdot \mv )m     
     - \lambda u\partial ^2[-Z_{10}m 
     \nonumber\\
     & +h] +  i\lambda g_5\mu^{\varepsilon/2} Z_{11}[\psi^{\dagger}\partial^2\psi-\psi\partial^2\psi^{\dagger}]\},
    \label{eq:ren_action}
\end{align}
where $\lambda, u, g_1, g_3, g_5, a_1, a_2$ are renormalized analogs of the bare parameters (written with the
subscript "0"), $Z_i,i=1,\ldots,11$ are renormalization
constants and
 $\mu$ is the renormalization mass \cite{Zinn,Vasiliev}.
A full specification of employed normalization conditions reads
\begin{align}
    \Gamma_{{\psi^{+}}'\psi'}  |_*  & = 2\lambda,\\
    \frac{\partial \Gamma_{m'm'} }{\partial k^2}
    \biggl|_* & = \lambda u,\\
    \frac{\partial \Gamma_{{\psi^{+}}' \psi} }{\partial (i\Omega)} \biggl|_* & = -\frac{1}{\lambda}
       \frac{\partial \Gamma_{{\psi^{+}}' \psi} }{\partial \mk^2} \biggl|_* = 1,
    \\
    \frac{\partial \Gamma_{m'm }}{\partial (i\Omega)} \biggl|_* & = - \frac{1}{\lambda u}
    \frac{\partial \Gamma_{m'm}}{\partial \mk^2} \biggl|_*  = 1,\\
    \Gamma_{ {\psi^{+}}'\psi m  } |_* & = -i\lambda g_3,\\
   \frac{\partial \Gamma_{m'\psi^+(\mpp)\psi(\mq)}}{\partial \mpp^2}
   \biggl|_*   
   & = - 
    \frac{\partial \Gamma_{m'\psi^+(\mpp)\psi(\mq)}}{\partial \mq^2}
   \biggl|_*  
   = i\lambda g_5,\\
   \Gamma_{ {\psi^+}' \psi^+ \psi \psi   } |_* & = -\frac{3\lambda g_1}{2},\\
   \frac{\partial \Gamma_{ {\psi^+}'\psi(\mpp) v_j(\mq) } }{\partial p_j} \biggl|_* & = 
   \frac{1}{a_1}\frac{\partial \Gamma_{ {\psi^+}'\psi(\mpp) v_j(\mq) } }{\partial q_j} \biggl|_* = -i,\\
   \frac{\partial \Gamma_{ m' m(\mpp) v_j(\mq) } }{\partial p_j} \biggl|_* & = 
   \frac{1}{a_2}\frac{\partial \Gamma_{ m' m(\mpp) v_j(\mq)} }{\partial q_j} \biggl|_* = -i.   
\end{align} 
For convenience, we have introduced $*$ coordinates specified as follows
\begin{equation}
  \Omega_i = 0, \quad k_i = 0, \quad \mu=l,
\end{equation}
where index $i$ enumerates independent external frequency or momenta entering given the 1PI function and
 the IR scale $l$ was introduced in Eq.~\eqref{eq:IRscale}.

The unrenormalized $\SA$ and the renormalized action functional $\SA_\text{R}$ are related by the 
standard formula 
\begin{equation*}
  \SA_\text{R}(\Phi)  = \SA(Z_{\Phi} \Phi).
\end{equation*}
Direct conseuqence of this formula are multiplicative relations for the fields 
$Z_{\Phi} \Phi = \{Z_\varphi \varphi: \varphi\in \Phi\}$,
and parameters~\cite{Vasiliev,Zinn}  
\begin{align}
    \lambda_0 & = \lambda Z_{\lambda},  &u_0&  = u Z_u, &g_{10}& = g_1 \mu^{\varepsilon} Z_{g_1},    
  \nonumber \\
    a_{10} & = a_1 Z_{a_1}, &a_{20}&  = a_2 Z_{a_2}, & w_0& = w \mu^{y} Z_w, 
  \nonumber \\
   g_{30} & = g_3 \mu^{\varepsilon/2} Z_{g_3}, &g_{50}& = g_5 \mu^{\varepsilon/2} Z_{g_5}.
  \label{eq:ren_param5}
\end{align}
Since the term ${\SA}_\text{vel}(\mv)$ given by (\ref{eq:action_vel}) is non-local in the spatial variable, we
know that according to general rules of the RG technique \cite{Vasiliev} it should not be renormalized. The
parameter $\alpha$ is not renormalized at all, i.e. $\alpha_0 = \alpha$, and serves as a free parameter of 
the theory. Due to the Galilean symmetry
 ensured by the presence of $\delta$-function in correlator~(\ref{eq:correlator}), both terms in the
 Lagrangian derivative $\nabla_t$ are renormalized with the same renormalization constants. 
 In addition, the quadratic term $ v D^{-1}v/2$  in the action~\eqref{eq:ren_action} is not renormalized because of a 
 passive nature of the advecting fields.
  As a direct consequence the velocity field $\mv$ is not renormalized and two relations follow
\begin{equation}
 Z_w Z_{\lambda} = 1, \quad Z_{\alpha} = Z_v = 1. 
\end{equation}

In the leading one-loop approximation, the 1-irreducible two-point Green functions take the form
\begin{align}
  \Gamma_{{\psi^{+}}'\psi'} & = 2\lambda Z_{1} +  
  \raisebox{-0.05cm}{\includegraphics[width=2cm]{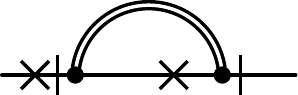}} +
  \raisebox{-0.05cm}{\includegraphics[width=2cm]{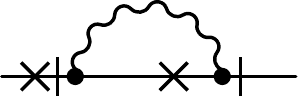}}, \\
  \Gamma_{m'm'} & = \lambda u k^{2} Z_{2} + 
  \raisebox{-0.05cm}{\includegraphics[width=2cm]{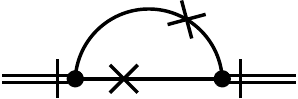}} +
  \raisebox{-0.05cm}{\includegraphics[width=2cm]{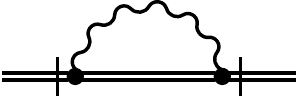}}, \\
  \Gamma_{{\psi^{+}}'\psi} & = i\Omega Z_{3} - \lambda \mk^{2} Z_{5} +  
  \raisebox{-0.05cm}{\includegraphics[width=2cm]{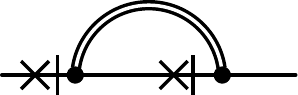}} \nonumber\\
  & + \raisebox{-0.05cm}{\includegraphics[width=2cm]{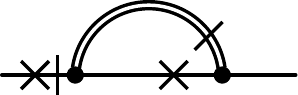}} 
   + \raisebox{-0.05cm}{\includegraphics[width=2cm]{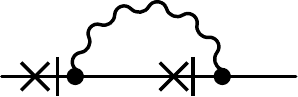}},
  \\
  \Gamma_{m'm} & = i\Omega Z_{8} - \lambda u \mk^{2} Z_{10} + 
  \raisebox{-0.05cm}{\includegraphics[width=2cm]{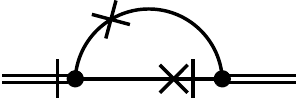}}
  \nonumber \\
  &+ \raisebox{-0.05cm}{\includegraphics[width=2cm]{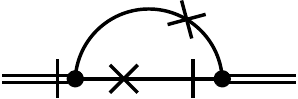}}
   + \raisebox{-0.05cm}{\includegraphics[width=2cm]{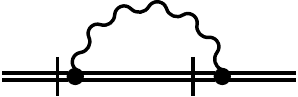}}.
\end{align}
Let us note that due to the structure of the vertex factors the relation $ Z_{m'}Z_{m} = 1 $ is fulfilled.

Further, the 1-irreducible Green functions 
accounting for
 non-linearities can be
 graphically represented as follows
\begin{align}
\Gamma_{{\psi^{+}}'\psi m} & =  - i \lambda g_{3} \mu^{\varepsilon/2} Z_{7} + 
\raisebox{-0.45cm}{\includegraphics[width=2.cm]{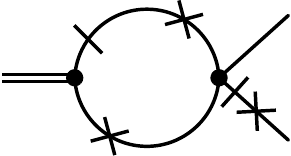}}
+ \raisebox{-0.4cm}{\includegraphics[width=1.8cm]{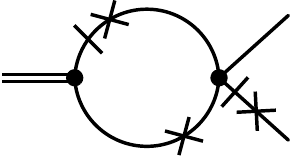}}
\nonumber \\ &  
+ \raisebox{-0.75cm}{\includegraphics[width=1.8cm]{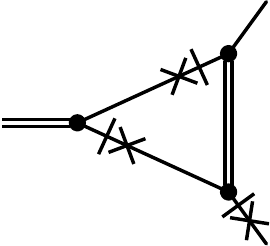}}
+ \raisebox{-0.75cm}{\includegraphics[width=1.8cm]{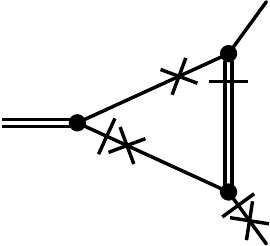}} 
+ \raisebox{-0.75cm}{\includegraphics[width=1.8cm]{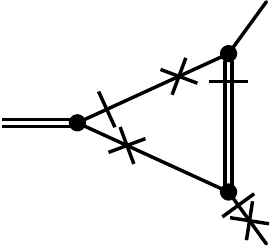}} 
\nonumber \\ &
+ \raisebox{-0.75cm}{\includegraphics[width=1.8cm]{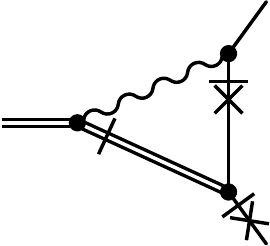}} 
+ \raisebox{-0.75cm}{\includegraphics[width=1.8cm]{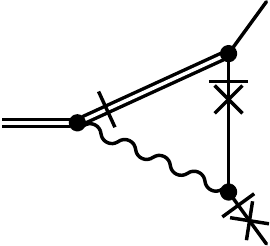}}  
+ \raisebox{-0.75cm}{\includegraphics[width=1.8cm]{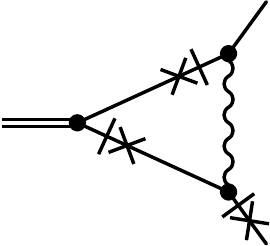}},\\
\Gamma_{m'\psi^{+}(\mpp)\psi(\mq)} & =  i \lambda g_{5} \mu^{\varepsilon/2} (\mpp^{2} - \mq^{2}) Z_{11} + 
   \raisebox{-0.4cm}{\includegraphics[width=1.8cm]{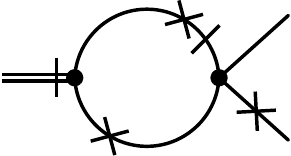}} 
\nonumber \\ & 
+ \raisebox{-0.4cm}{\includegraphics[width=1.8cm]{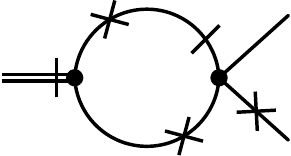}}    
+ \raisebox{-0.75cm}{\includegraphics[width=1.8cm]{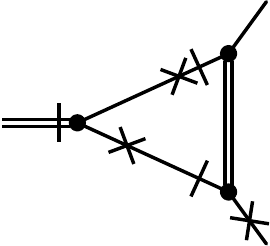}}
+ \raisebox{-0.75cm}{\includegraphics[width=1.8cm]{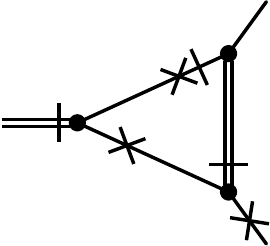}} 
\nonumber \\ & 
+ \raisebox{-0.75cm}{\includegraphics[width=1.8cm]{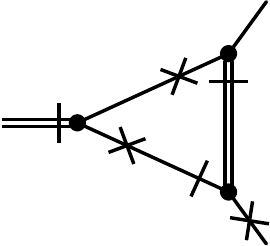}} 
+ \raisebox{-0.75cm}{\includegraphics[width=1.8cm]{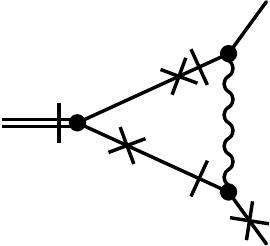}}
+ \raisebox{-0.75cm}{\includegraphics[width=1.8cm]{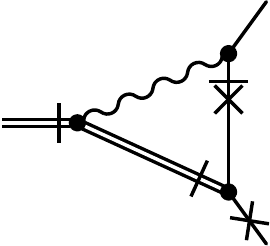}} 
\nonumber \\ & 
+ \raisebox{-0.75cm}{\includegraphics[width=1.8cm]{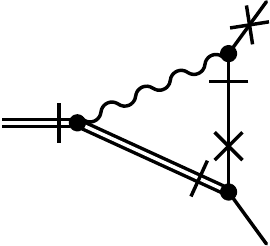}}, \\
\Gamma_{{\psi^{+}}'\psi^{+}\psi\psi} & =\ - \frac{2}{3} \lambda g_{1} \mu^{\varepsilon} Z_{6} 
+ \raisebox{-0.4cm}{\includegraphics[width=1.8cm]{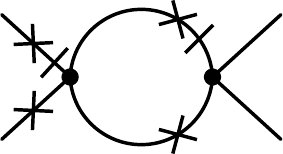}}  
+ \raisebox{-0.4cm}{\includegraphics[width=1.8cm]{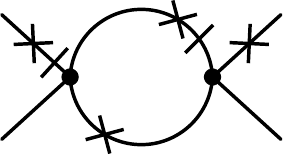}} 
\nonumber \\ &
+ \raisebox{-0.4cm}{\includegraphics[width=1.8cm]{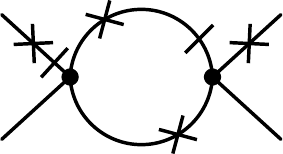}} 
+ \raisebox{-0.75cm}{\includegraphics[width=1.8cm]{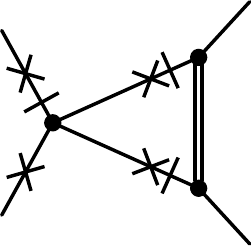}}
+ \raisebox{-0.75cm}{\includegraphics[width=1.8cm]{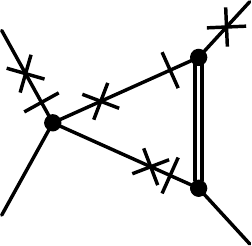}} 
\nonumber \\ &
+ \raisebox{-0.75cm}{\includegraphics[width=1.8cm]{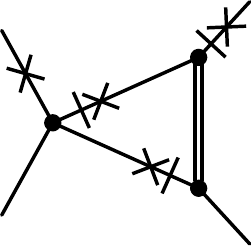}}
+ \raisebox{-0.75cm}{\includegraphics[width=1.8cm]{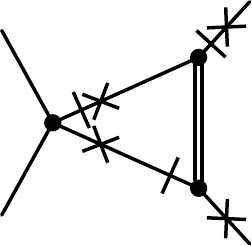}}
+ \raisebox{-0.75cm}{\includegraphics[width=1.8cm]{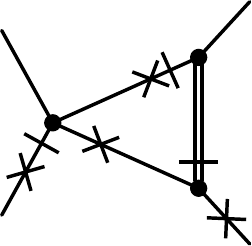}}
\nonumber \\ &
+ \raisebox{-0.75cm}{\includegraphics[width=1.8cm]{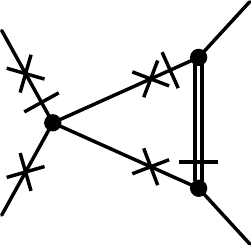}} 
+ \raisebox{-0.75cm}{\includegraphics[width=1.8cm]{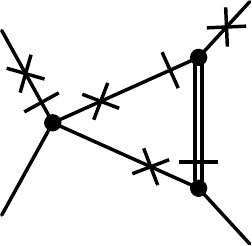}} 
+ \raisebox{-0.75cm}{\includegraphics[width=1.8cm]{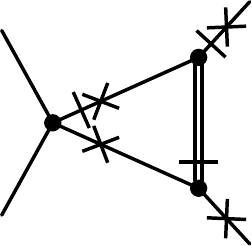}} 
\nonumber \\ &
+ \raisebox{-0.75cm}{\includegraphics[width=1.8cm]{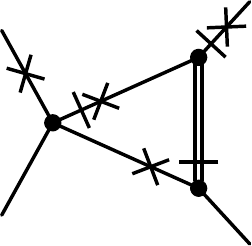}} 
+ \raisebox{-0.75cm}{\includegraphics[width=1.8cm]{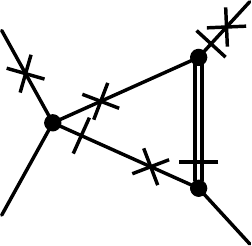}} 
+ \raisebox{-0.75cm}{\includegraphics[width=1.8cm]{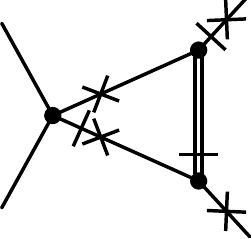}} 
\nonumber \\ & 
+ \raisebox{-0.9cm}{\includegraphics[width=1.8cm]{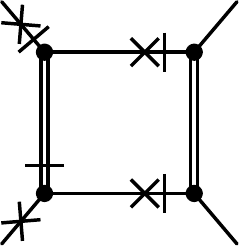}} 
+ \raisebox{-0.9cm}{\includegraphics[width=1.8cm]{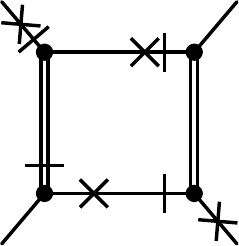}}  
+ \raisebox{-0.9cm}{\includegraphics[width=1.8cm]{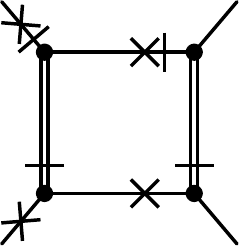}} 
\nonumber \\ & 
+ \raisebox{-0.9cm}{\includegraphics[width=1.8cm]{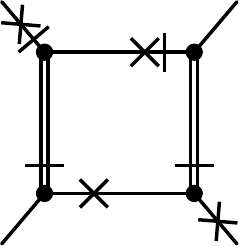}}  
+ \raisebox{-0.9cm}{\includegraphics[width=1.8cm]{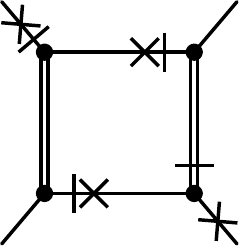}} 
+ \raisebox{-0.9cm}{\includegraphics[width=1.8cm]{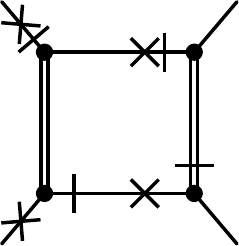}} 
\nonumber \\ & 
+ \raisebox{-0.9cm}{\includegraphics[width=1.8cm]{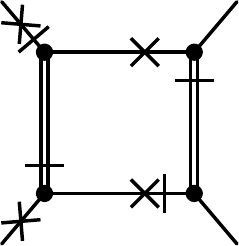}} 
+ \raisebox{-0.9cm}{\includegraphics[width=1.8cm]{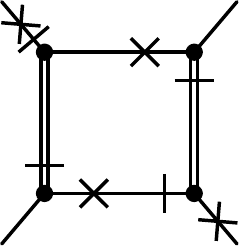}} 
+ \raisebox{-0.9cm}{\includegraphics[width=1.8cm]{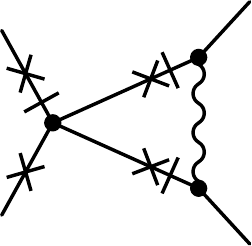}} 
\nonumber \\ & 
+ \raisebox{-0.9cm}{\includegraphics[width=1.8cm]{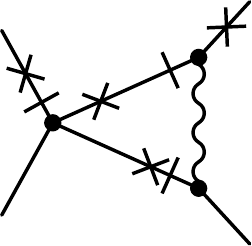}} 
+ \raisebox{-0.9cm}{\includegraphics[width=1.8cm]{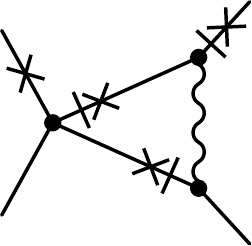}}  
+ \raisebox{-0.9cm}{\includegraphics[width=1.8cm]{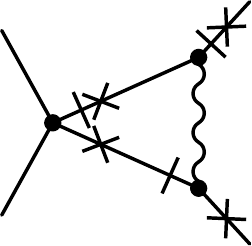}} 
\nonumber \\ & 
+\raisebox{-0.9cm}{\includegraphics[width=1.8cm]{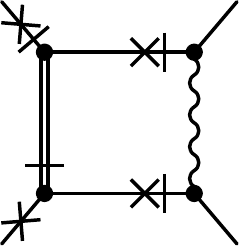}} 
+ \raisebox{-0.9cm}{\includegraphics[width=1.8cm]{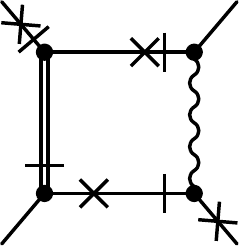}}  
+ \raisebox{-0.9cm}{\includegraphics[width=1.8cm]{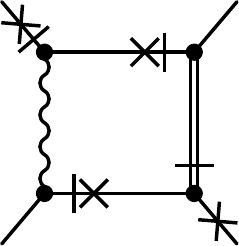}} 
\nonumber \\ & 
+ \raisebox{-0.9cm}{\includegraphics[width=1.8cm]{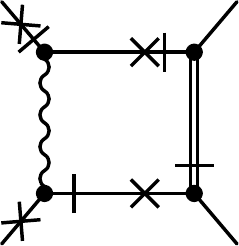}},\\
\Gamma_{\psi^{+'} \psi(\mpp) v_j(\mq)} & =\ -i p_j Z_3 - i a_1 q_j Z_4 
+ \raisebox{-0.4cm}{\includegraphics[width=1.8cm]{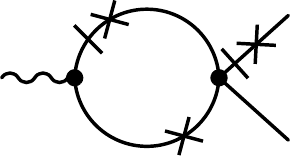}}
\nonumber \\ &
+ \raisebox{-0.4cm}{\includegraphics[width=1.8cm]{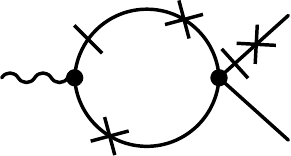}} 
+ \raisebox{-0.75cm}{\includegraphics[width=1.8cm]{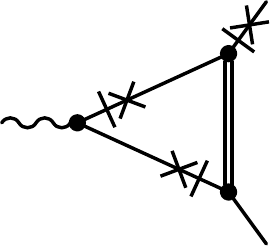}}
+ \raisebox{-0.75cm}{\includegraphics[width=1.8cm]{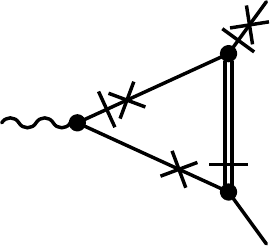}}
\nonumber \\ &
+ \raisebox{-0.75cm}{\includegraphics[width=1.8cm]{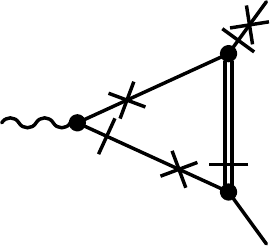}}
+ \raisebox{-0.75cm}{\includegraphics[width=1.8cm]{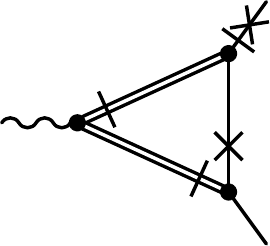}}
+ \raisebox{-0.75cm}{\includegraphics[width=1.8cm]{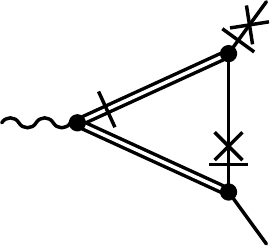}}
\nonumber \\ &
+ \raisebox{-0.75cm}{\includegraphics[width=1.8cm]{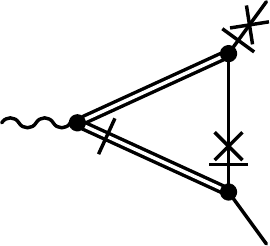}}
+ \raisebox{-0.75cm}{\includegraphics[width=1.8cm]{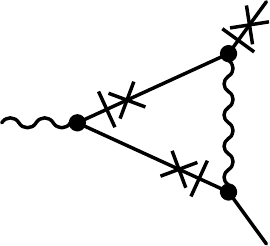}},\\
\Gamma_{m^{'} m(\mpp) v_j(\mq)} &=\ -i p_j Z_8 - i a_2 q_j Z_9 
+ \raisebox{-0.75cm}{\includegraphics[width=1.8cm]{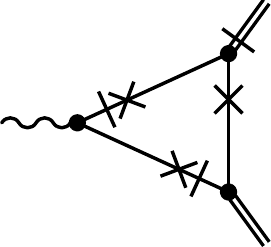}}
\nonumber \\ & 
+ \raisebox{-0.75cm}{\includegraphics[width=1.8cm]{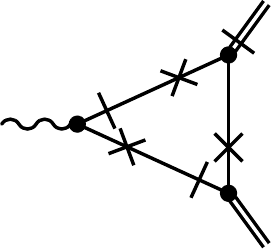}}
+ \raisebox{-0.75cm}{\includegraphics[width=1.8cm]{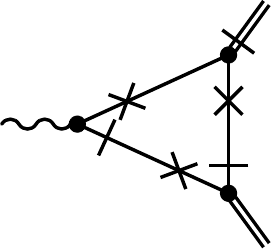}}
+ \raisebox{-0.75cm}{\includegraphics[width=1.8cm]{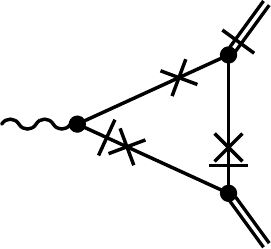}}
\nonumber \\ &
+ \raisebox{-0.75cm}{\includegraphics[width=1.8cm]{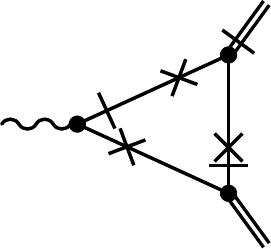}}
+ \raisebox{-0.75cm}{\includegraphics[width=1.8cm]{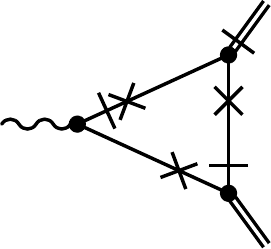}}
+ \raisebox{-0.75cm}{\includegraphics[width=1.8cm]{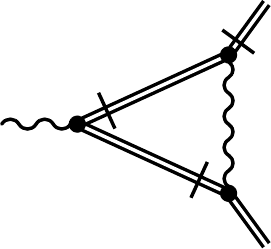}}
.
\end{align}

A major reduction of divergent diagrams comes from two observations. First, according
to general rules of critical dynamics~\cite{Vasiliev} any Feynman graph constructed
solely from retarded propagators does not possess UV divergence. Second, a structure of
interaction vertices allows, in some cases, to pull out momentum dependence and thus reduce
 the effective dimension of internal momentum integration. This then also leads to UV convergence
of the corresponding Feynman diagram.

 A calculation in the employed RG scheme proceeds in a standard fashion and
we have summarized all the results in Appendix~\ref{sec:app_feynman}. Note that
prefactors contain explicit $d$-dependence stemming from vector and tensorial
character of interactions. 

Technical difficulties related to the chosen IR cutoff were circumvented by a proper extraction of external momentum
from a given diagram. Once correct frequency or momentum dependence was pulled out of a diagram, it was permissible to set all
external frequencies and momenta inside integral to zero. This greatly simplifies calculations of divergent parts of Feynman diagrams.
However, we expect this to be much more cumbersome in two- and higher-loop approximations.
{\subsection{RG functions} \label{sec:rg_func}}

RG invariance \cite{Vasiliev,Zinn} can be conveniently expressed by the
differential equation  $\DA_{\mu} \Gamma = 0$, where the differentiation with
respect to renormalization mass $\mu$ in the operator $\DA_{\mu} = \mu \partial_{\mu}$ is performed at fixed values of
 (bare) unrenormalized variables. For the 1PI renormalized Green function, it can be rewritten as 
$[\DA_{RG} - n_{\Phi}\gamma_{\Phi}] \Gamma_R (\mu, \dots ) = 0$, where $n_{\Phi}$ is the number of
fields of a given renormalized 1PI Green function. The operator $\DA_{RG}$ stands for $\DA_{\mu}$ in terms of
the renormalized variables
\begin{equation}
  \DA_{RG}\equiv \mu \frac{\partial}{\partial \mu}\biggl|_0 = \mu \frac{\partial}{\partial \mu} + 
  \sum_{g_i \in g} \beta_{g_i}\frac{\partial}{\partial g_i} -  \gamma_\lambda \lambda \frac{\partial}{\partial \lambda},
  \label{eq:diff_oper}
\end{equation}
where the  summation runs over all charges of the theory. For convenience we have introduced compact
  notation 
\begin{equation}
  g \equiv \{ g_1, g_3, g_5, u, w, a_1, a_2 \}.
  \label{eq:def_charges}
\end{equation}
The differentiation in~\eqref{eq:diff_oper} at fixed values of the bare parameters is indicated explicitly by the subscript "0". 
 The beta functions $\beta_{g_i}$ and anomalous dimensions $\gamma_F$ of the mode are defined by the logarithmic
derivatives \cite{Vasiliev,Zinn}
\begin{equation}
  \beta_i =  \frac{\partial g_i}{\partial\ln \mu}\biggl|_0, \quad \gamma_F = \frac{\partial \ln Z_F}{\partial \ln \mu}\biggl|_0.
  \label{eq:RG_func}
\end{equation}
straightforward application of these definitions onto the relations (\ref{eq:ren_param5}) yields
\begin{align}
  \beta_{g_1} & = g_1(-\varepsilon - \gamma_{g_1}), &\beta_{g_3}& = g_3\left(-\frac{\varepsilon}{2} - \gamma_{g_3}\right), \\
  \beta_{g_5} & = g_5\left(-\frac{\varepsilon}{2} - \gamma_{g_5}\right), &\beta_w& = w(-y - \gamma_w), \\
  \beta_{a_1} & = - a_1 \gamma_{a_1}, &\beta_{a_2} &= - a_2 \gamma_{a_2},\\
  \beta_u & = - u \gamma_u, &\beta_{\alpha}& = - \alpha \gamma_{\alpha}.
  \label{eq:beta_func}
\end{align}
In particular, the $\beta_{\alpha}$ function identically vanishes due to the aforementioned constraint $Z_\alpha=1$. 
To obtain the remaining renormalization 
constants in the MS scheme, the UV-divergent terms (poles in $\varepsilon$ and $y$ in our case) have to be extracted from
the loop expansion
of the corresponding 1PI functions. Renormalization constants $Z_i,i=1,\ldots,11$
 are related to the renormalization constants of the parameters and fields by  means of relations
\begin{align} 
   Z_{\lambda} & = Z_w^{-1} = \frac{Z_5}{ Z_3}, &Z_u& = \frac{Z_{10} Z_3}{ Z_5 Z_8}, 
   \label{eq:rg_constants1} \\
   Z_{g_5} & = \frac{ Z_{11} Z_1}{Z_5^2} \left(  \frac{Z_{10}}{Z_{2} Z_8} \right)^{1/2},
   &Z_{g_1} & = \frac{Z_6 Z_1}{ Z_5^{2} Z_3},
   \label{eq:rg_constants2}
   \\   
   Z_{g_3}& = \frac{Z_7}{Z_5} \biggl( \frac{ Z_2 }{ Z_{10} Z_8 }\biggl)^{1/2}, &Z_{a_1}& = \frac{Z_4}{ Z_3} ,
   \label{eq:rg_constants3}
   \\
   Z_{\psi^\prime}&= \biggl( \frac{Z_1 Z_3}{ Z_5}\biggl)^{1/2},  &Z_{\psi}&  = \biggl( \frac{Z_3 Z_5}{ Z_1} \biggl)^{1/2},   
   \label{eq:rg_constants4}   \\
   Z_m & =   \left( \frac{Z_{10} Z_8 }{ Z_2 }\right)^{1/2},
   &Z_{a_2}& = \frac{Z_9}{ Z_8},
   \label{eq:rg_constants5}\\
   Z_{m'} & = \left( \frac{Z_{2} Z_8 }{ Z_{10} }\right)^{1/2},
   \label{eq:rg_constants6}\\
\end{align}
 where we have used two additional relations \cite{Vasiliev} for fields renormalization
\begin{equation}
  Z_{\psi^\prime} = Z_{{\psi^{\dagger}}^\prime}, \quad Z_{\psi^\dagger} = Z_\psi.  
\end{equation}
From the second relation in (\ref{eq:RG_func}) anomalous dimensions $\gamma_F$  can be directly obtained from 
 the renormalization constants (\ref{eq:rg_constants1})-(\ref{eq:rg_constants5}). 
 A special feature of the one-loop approximation is the fact that to this order we have found 
\begin{equation}
  Z_8 = Z_9 = 1.
   \label{eq:z8z9}
\end{equation}
Substituting \eqref{eq:z8z9} in~\eqref{eq:rg_constants5} leads to
\begin{equation}
  Z_{a_2} =1.
\end{equation}
Hence, parameter $a_2$ can be also regarded as a free parameter of the model to the order of perturbation theory.
{\section{Scaling regimes and the fixed-point structure} \label{sec:regimes}}
From an experimental point of view, most relevant for statistical physics is the IR-asymptotic behavior, i.e. behavior
of Green functions at small frequencies $\omega\rightarrow 0$ and momenta $\mk \rightarrow 0$. This is related to
large-scale macroscopic regimes of a given renormalizable field theoretic model, which are associated with IR attractive 
fixed  points of the corresponding 
RG equations \cite{Vasiliev,Zinn}.  A fixed point (FP) is defined as  such point 
$g^{*} \equiv \{ g_1^{*}, g_3^{*}, g_5^{*}, u^{*}, a_1^{*}, a_2^{*}, w^{*} \}$ for which all $\beta$ functions
simultaneously vanish, i.e. 
\begin{equation}
  \beta_{g_i}(g^{*}) = 0,
  \label{eq:zero_beta}
\end{equation}
where $g_i$ is any member of the set $g$ defined in Eq.~\eqref{eq:def_charges}. 
 The IR stability of a given fixed point is determined by the matrix of first derivatives of $\beta$ functions
\begin{equation}
  \Omega_{ij} = \frac{\partial \beta_i}{\partial g_j}\biggl|_{g^{*}}. 
  \label{eq:omega}
\end{equation}
Coordinates of fixed points do not possess a direct physical information, because they depend on the chosen renormalization
scheme \cite{Vasiliev,Zinn}. However, in perturbation theory 
  universal quantities and number of fixed points actually depend on the chosen RG scheme~\cite{honkonen_chaos}.
 In order to proceed in actual calculations, we have expanded prefactors containing $d$-parameter in
 Feynman diagrams  by $d=4-\varepsilon$ (see Appendix~\ref{sec:app_feynman} for details). Though not correct in higher loop calculations, in the one-loop
 approximation this can be regarded as an adequate operation. 
 
For
completeness, we list all found 
 fixed points in Appendix~\ref{app:coord}.

 For  IR attractive fixed points all real parts of eigenvalues of matrix (\ref{eq:omega}) are positive.
Moreover, physical conditions $u^{*} >0$ and $w^{*} >0$ have to be fulfilled, which is due to their appearance in free pair 
correlation functions 
(see Eq.~(\ref{eq:propa1}) and Eq.~(\ref{eq:propa6})). 
 
A thorough analysis
of beta functions has revealed that there are several possible regimes in case without thermal 
fluctuations, i.e., regimes with a fixed point coordinate $g^*_3 = 0$. It is worth mentioning
  that for a purely transverse velocity field ($\alpha=0$) the terms containing 
$a_{10}$, $a_{20}$ in (\ref{eq:replace1})-(\ref{eq:replace2}) and all subsequent expressions vanish, and such parameters
disappear from the model. 

An actual analysis proved to be rather cumbersome and not very illuminating. The main technical problems were related to an 
appearance of 
parameters $a_1,a_2$ brought about by compressibility of velocity field in action~\eqref{eq:ren_action}. Moreover,
compressibility parameter $\alpha$ is 
free from any restriction and can attain any positive value~\cite{Ant00}. Thus when possible, we therefore try to present
our findings by graphical means.
Technical details can be found in the Appendices.

Altogether eight fixed points have been found. However, only some of them are IR stable.
A trivial Gaussian-like fixed point FP1 is IR-stable in the region restricted by inequalities
\begin{equation}
  \varepsilon<0,\quad y<0.
  \label{eq:FP1}
\end{equation}
FP1 corresponds to a free model, for which all interactions are irrelevant and is stable above upper critical space 
dimension $d>d_c=4$ . The corresponding 
critical exponents attain their mean-field values.
 
The fixed points FP2, FP3 and FP4 are unstable because the corresponding eigenvalues of the $\Omega$ matrix~\eqref{eq:omega} 
(see Tab.~\ref{tab:eigenvalues1}) 
always contain one positive and one negative eigenvalue for any value of exponent $\eps.$

\begin{figure}[!tbp]
  \centering
  \subfloat{\includegraphics[width=0.4\textwidth]{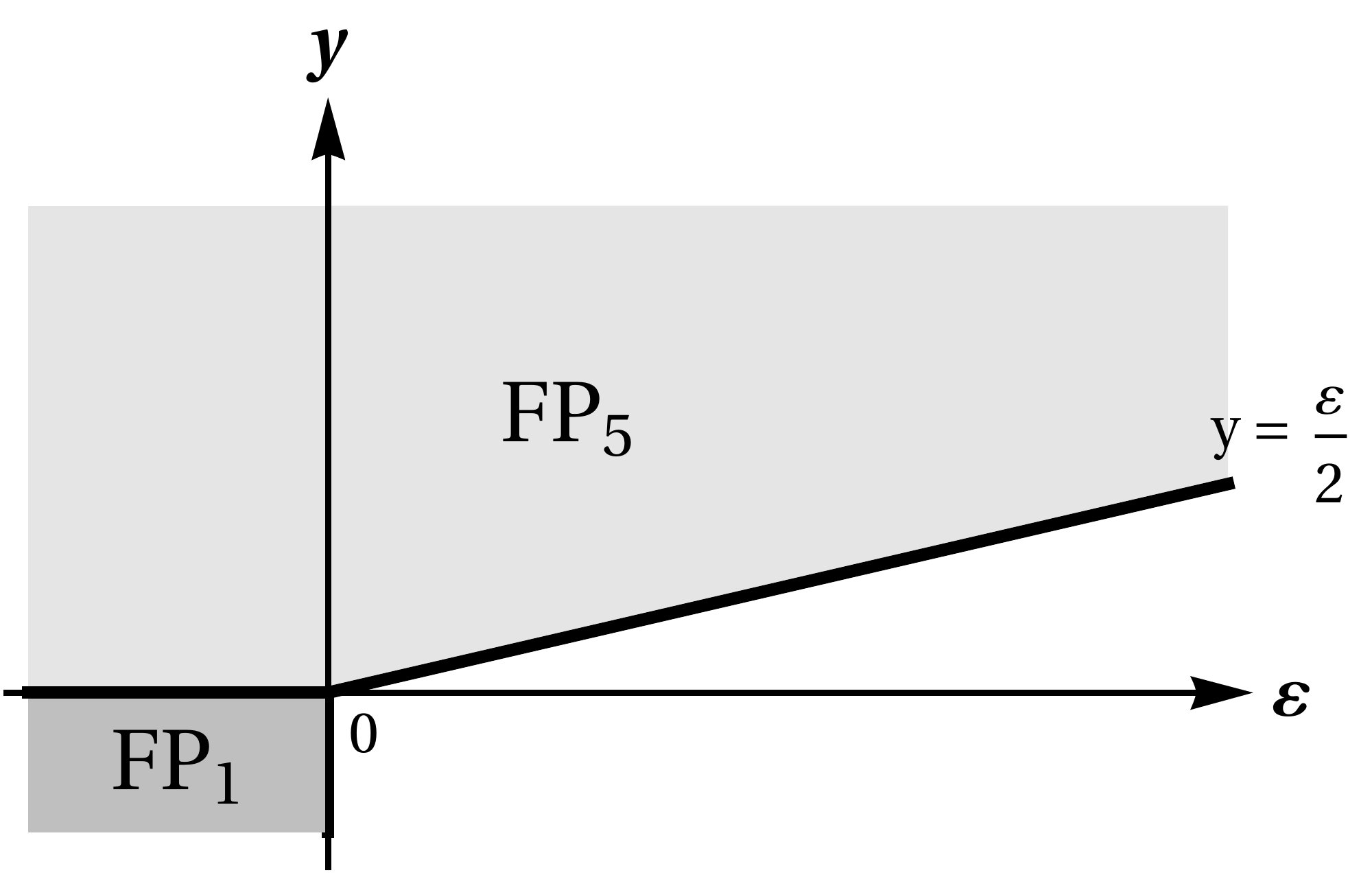}\label{fig_a}}
  \hfill
  \subfloat{\includegraphics[width=0.4\textwidth]{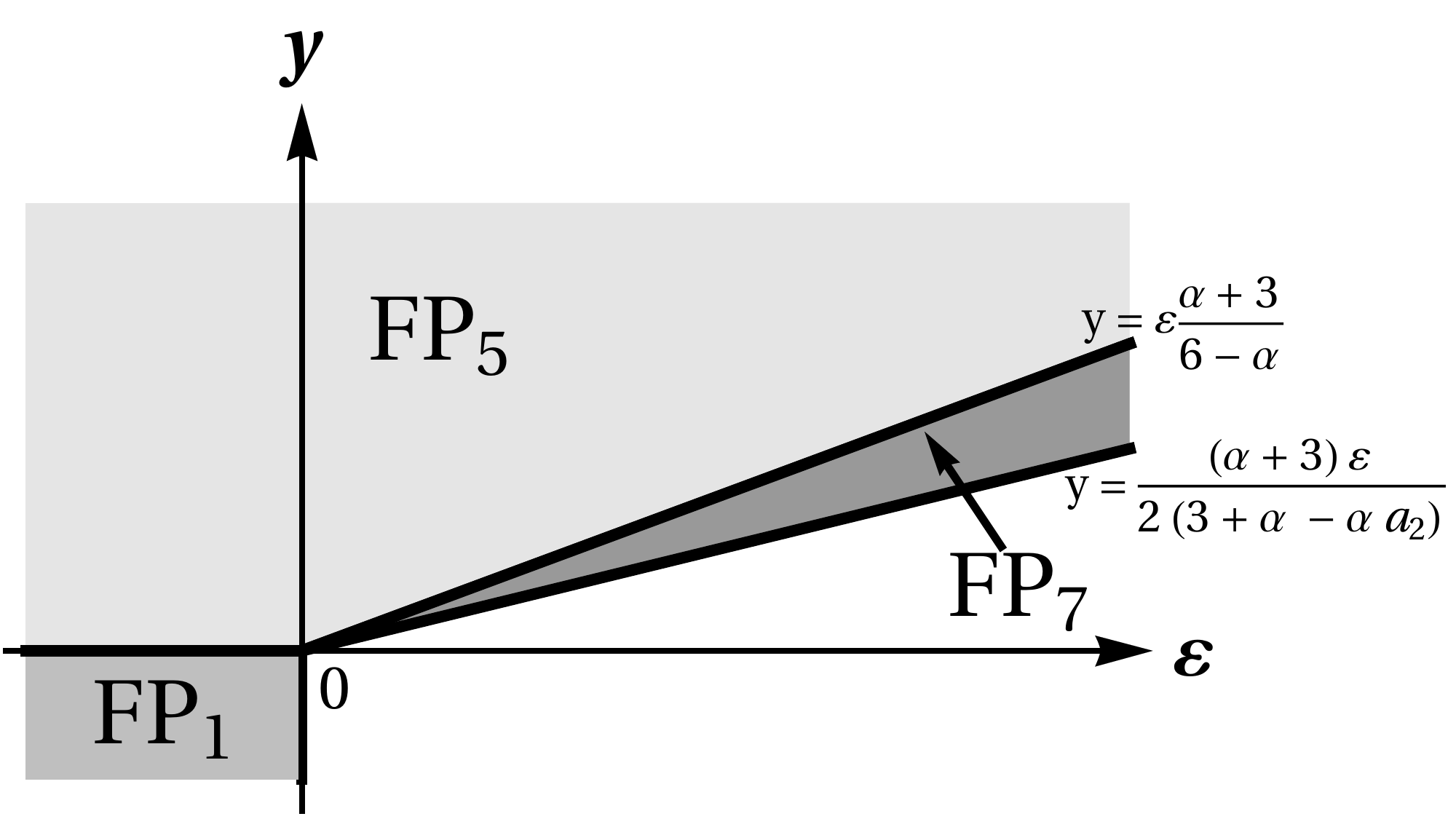}\label{fig_b}}
  \caption{Regions of stability of the fixed points in the model
   for compressibility parameter $\alpha=0$
   (top) and for the following restricted choice of parameters $a_1 = {1}/{4}$, $\alpha \in (0, \frac{8}{5})$ and $a_2 \in (0, \frac{3}{2})$  (bottom).  
  }
  \label{fig:graph1}
\end{figure}

\begin{figure}[!tb]
  \centering
  \subfloat{\includegraphics[width=0.4\textwidth]{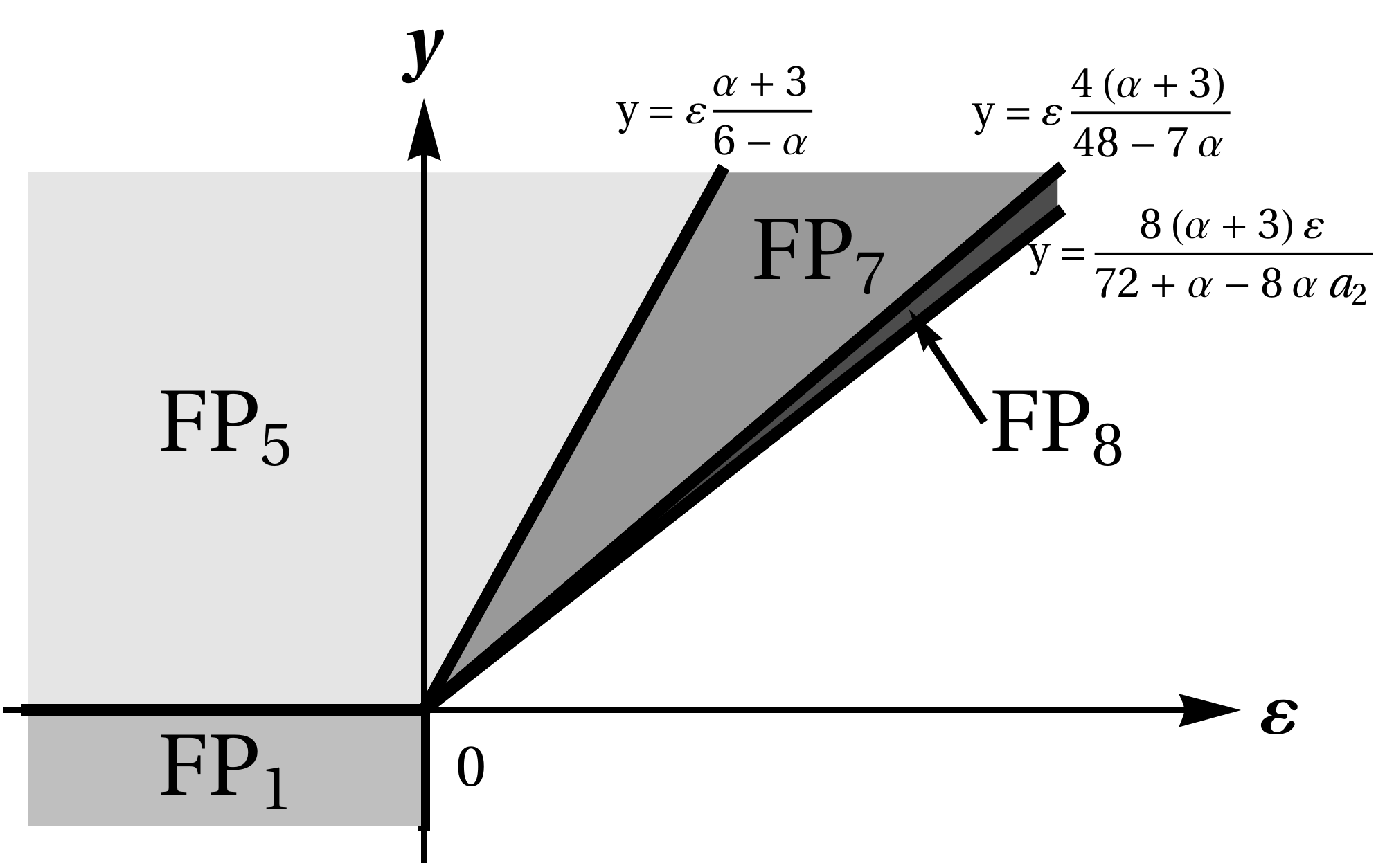}\label{fig_c}}
  \hfill
  \subfloat{\includegraphics[width=0.4\textwidth]{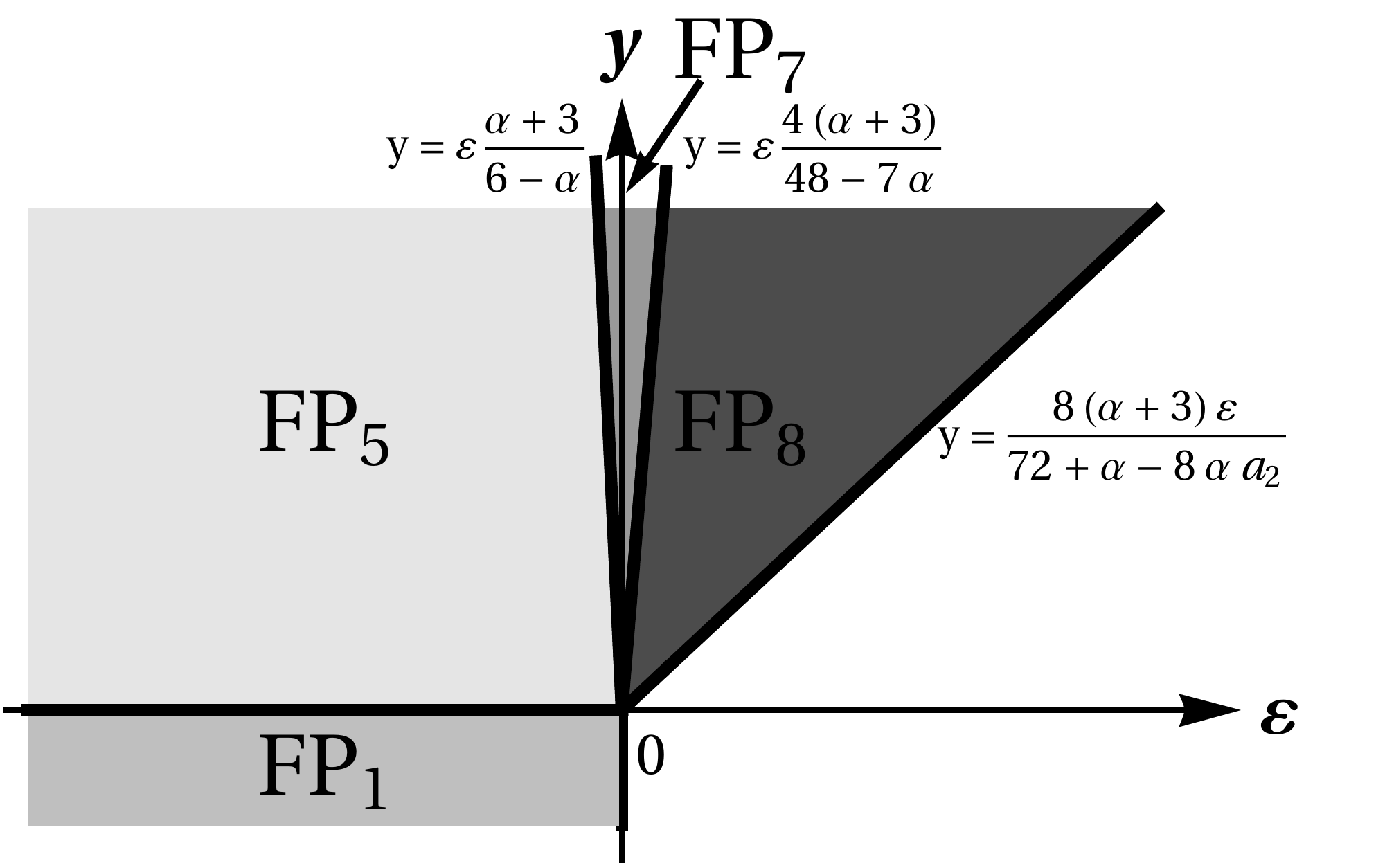}\label{fig_d}}
  \caption{Stability regions of the fixed points for which $a_1 = {1}/{4}$
  and for the following restricted choice of parameters
   $\alpha \in (\frac{8}{5}, 6)$, $a_2 \in (0, \frac{15}{8}-\frac{3}{\alpha})$
  (top)  
  and  $\alpha \in (6,\frac{48}{7})$, $a_2 \in (0, \frac{15}{8}-\frac{3}{\alpha})$
  (bottom).
  }
  \label{fig:graph2}
\end{figure}

\begin{figure}[!tb]
  \centering
  \subfloat
  {\includegraphics[width=0.4\textwidth]{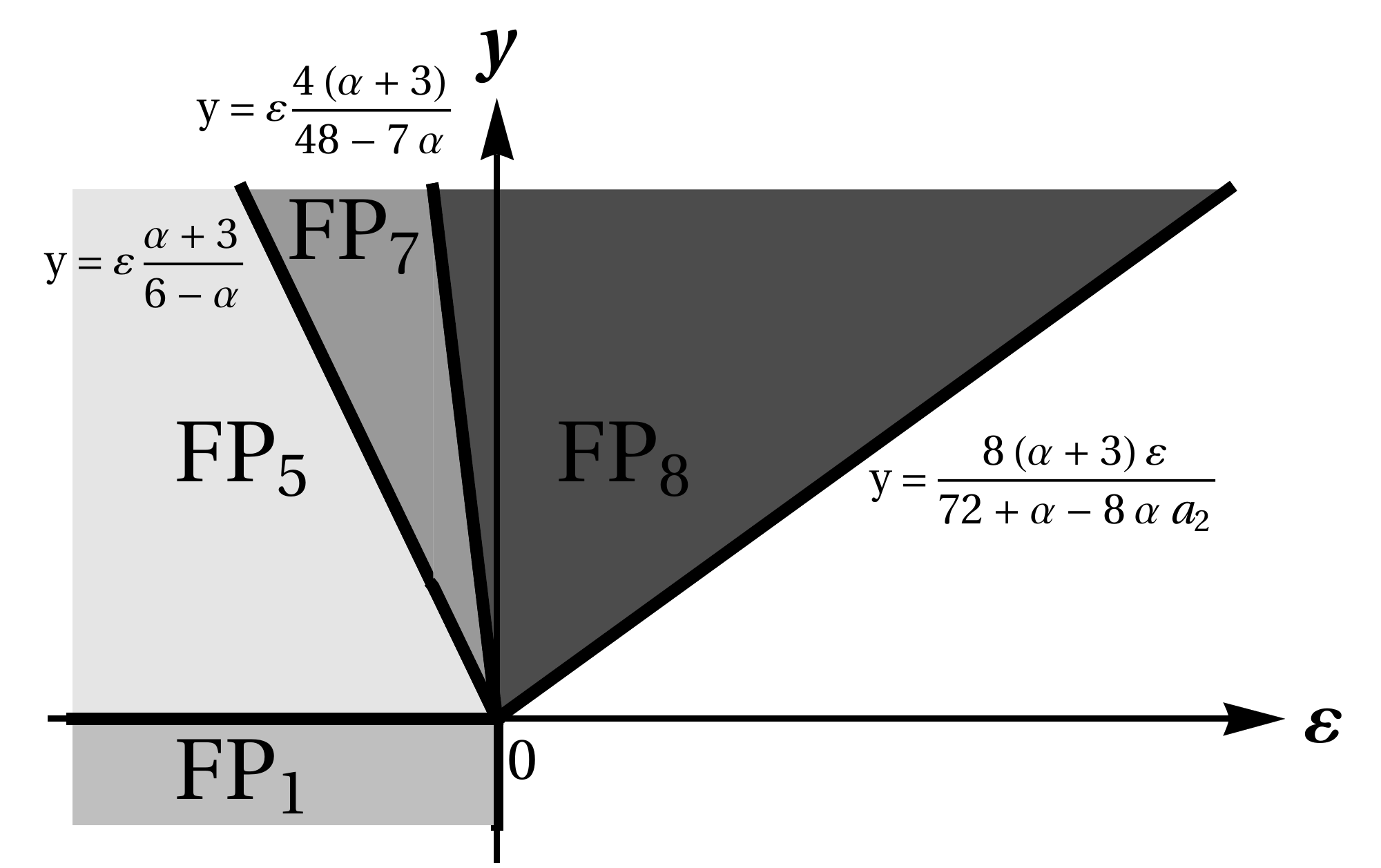}\label{fig_e}}
  \hfill
  \subfloat
  {\includegraphics[width=0.4\textwidth]{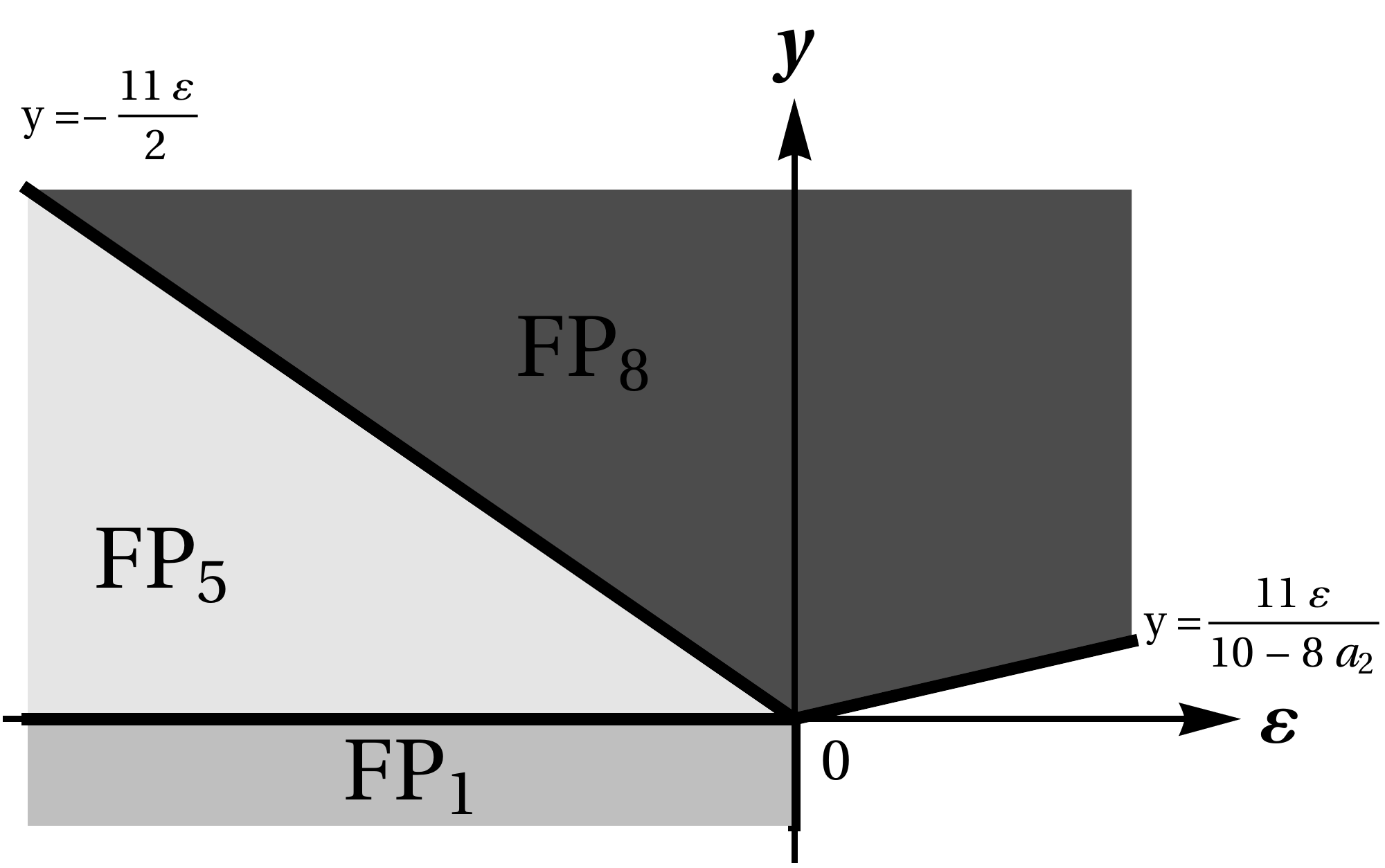}\label{fig_f}}
  \caption{Stability regions of the fixed points for which $a_1 = {1}/{4}$
  and for the following restricted choice of parameters
    $\alpha \in (\frac{48}{7}, 8)$, $ a_2 \in (0, \frac{1}{8}+\frac{9}{\alpha})$
  (top)
  and $\alpha = 8,a_2 \in (0, \frac{5}{4})$
  (bottom).  
  }
  \label{fig:graph3}
\end{figure}

For FP5 fixed points' values of charges $g_1,g_3$ and $g_5$ vanish identically. The nonzero coordinates $w$ and $u$ hint
at IR relevance of the turbulent advection. 
This regime corresponds to a
well-known passive advection problem \cite{Ant00,FGV01}. We recall that parameters $\alpha$, $a_1$ and $a_2$ should be
regarded as free parameters  
 and stability regions might exhibit a non-trivial dependence on them. This point is IR-stable in the region restricted
 effectively by two inequalities. 
 First one is simply the condition $y>0$ and the second restriction is given by one of the following inequalities,
\begin{align}
 [3 - \alpha( 3 a_1^2-3a_1+1)] y &> \varepsilon \frac{3+\alpha}{4},
\label{eq:FP5_a} \\
 [3 - \alpha( 8 a_1^2-4a_1+1)] y &> \varepsilon \frac{3+\alpha}{2},
\label{eq:FP5_b} \\
 [3 - \alpha(4 a_1 a_2 - 1)] y &> \varepsilon \frac{3+\alpha}{2}.
\label{eq:FP5_c}
\end{align}
Which one of them will be imposed depends on the value of free parameters $\alpha$, $a_1$ and $a_2$. 
The corresponding inequality for arbitrary value of parameters can be found in 
Tab.~\ref{tab:fp5_inequality}.  Parameter values corresponding to the endpoints in Tab.~\ref{tab:fp5_inequality} lead 
to inequalities 
 of the same form. A thorough analysis of FP5 reveals some interesting features. Increasing one of the free parameters with two 
 remaining parameters fixed, boundaries of stability region shift. From a technical point of view, this is caused by a 
 form of the left hand side of inequalities~\eqref{eq:FP5_a}-\eqref{eq:FP5_c}. For instance, we choose
 $a_1 = 1/4$ and $a_2$ from the interval $\left(0, \frac{3}{2}\right)$. Then, whenever the compressibility
 parameter $\alpha$ attains a value smaller than $\alpha^{*} = 8$, the region of
stability (Fig.~\ref{fig_b}, \ref{fig:graph2}, \ref{fig_e}) is restricted by inequality (\ref{eq:FP5_b}). On the other hand, for
 $\alpha$ larger than $\alpha^{*}$, restrictions comes from inequality (\ref{eq:FP5_a}). 
  For a special case $\alpha^{*} = 8$, the inequalities (\ref{eq:FP5_a}) and (\ref{eq:FP5_b}) are actually of the same
  form (Fig.~\ref{fig_f}).  
 For $\alpha=6$ the inequality (\ref{eq:FP5_b}) takes a simple form of $\varepsilon<0$.  With an increasing value of $\alpha$, the
 boundary of FP5 rotates in the 
counterclockwise direction, and in the purely potential limit $\alpha \rightarrow \infty$ it approaches the ray
$ y = -\frac{4}{7}\varepsilon$. Let 
us note that in a one-loop approximation boundaries of stability regions between different fixed points are often given 
by straight lines. In a higher-loop
 calculations overlapping regions might appear.

\begin{table}[!ht]
\renewcommand{\arraystretch}{1.75}
\begin{tabular}{|c|c|c|c|}
\hline
$a_1$ & $\alpha$ & $a_2$ & Ineq. \\
\hline
\hline
$\left(0, \frac{\sqrt{3}-1}{2}\right)$ & $\left(0,\alpha^{*}\right)$ & $\left(0, \frac{4a_1^2-2a_1+1}{2a_1}\right)$ & (\ref{eq:FP5_b}) \\
\hline
$\left(0, \frac{\sqrt{3}-1}{2}\right)$ & $\left(0,\alpha^{*}\right)$ & $\left( \frac{4a_1^2-2a_1+1}{2a_1}, \infty \right)$ & (\ref{eq:FP5_c}) \\
\hline
$\left(0, \frac{\sqrt{3}-1}{2}\right)$ & $\left(\alpha^{*}, \infty \right)$ & $\left(0, \frac{3\alpha(2a_1^2-2a_1+1)-3}{4 a_1\alpha}\right)$ & (\ref{eq:FP5_a})\\
\hline
$\left(0, \frac{\sqrt{3}-1}{2}\right)$ & $\left(\alpha^{*}, \infty \right)$ & $\left(\frac{3\alpha(2a_1^2-2a_1+1)-3}{4 a_1\alpha}, \infty\right)$ & (\ref{eq:FP5_c})\\
\hline
$\left(\frac{\sqrt{3}-1}{2}, \infty\right)$ & $\left(0,\infty \right)$ & $\left(0, \frac{4a_1^2-2a_1+1}{2a_1}\right)$ & (\ref{eq:FP5_b}) \\
\hline
$\left(\frac{\sqrt{3}-1}{2}, \infty\right)$ & $\left(0,\infty \right)$ & $\left(\frac{4a_1^2-2a_1+1}{2a_1}, \infty\right)$ & (\ref{eq:FP5_c}) \\
\hline
\end{tabular}
\caption{ Intervals of free parameters $a_1$, $\alpha$, $a_2$ with the corresponding inequality 
 restricting the stability region of FP5. Here, $\alpha^{*} $ stands for the expression 
 $ {3}/(1-2a_1-2 a_1^2)$ .
 }
\label{tab:fp5_inequality}
\end{table}

\begin{figure}[!tbp]
  \centering
  \subfloat
  {\includegraphics[width=0.4\textwidth]{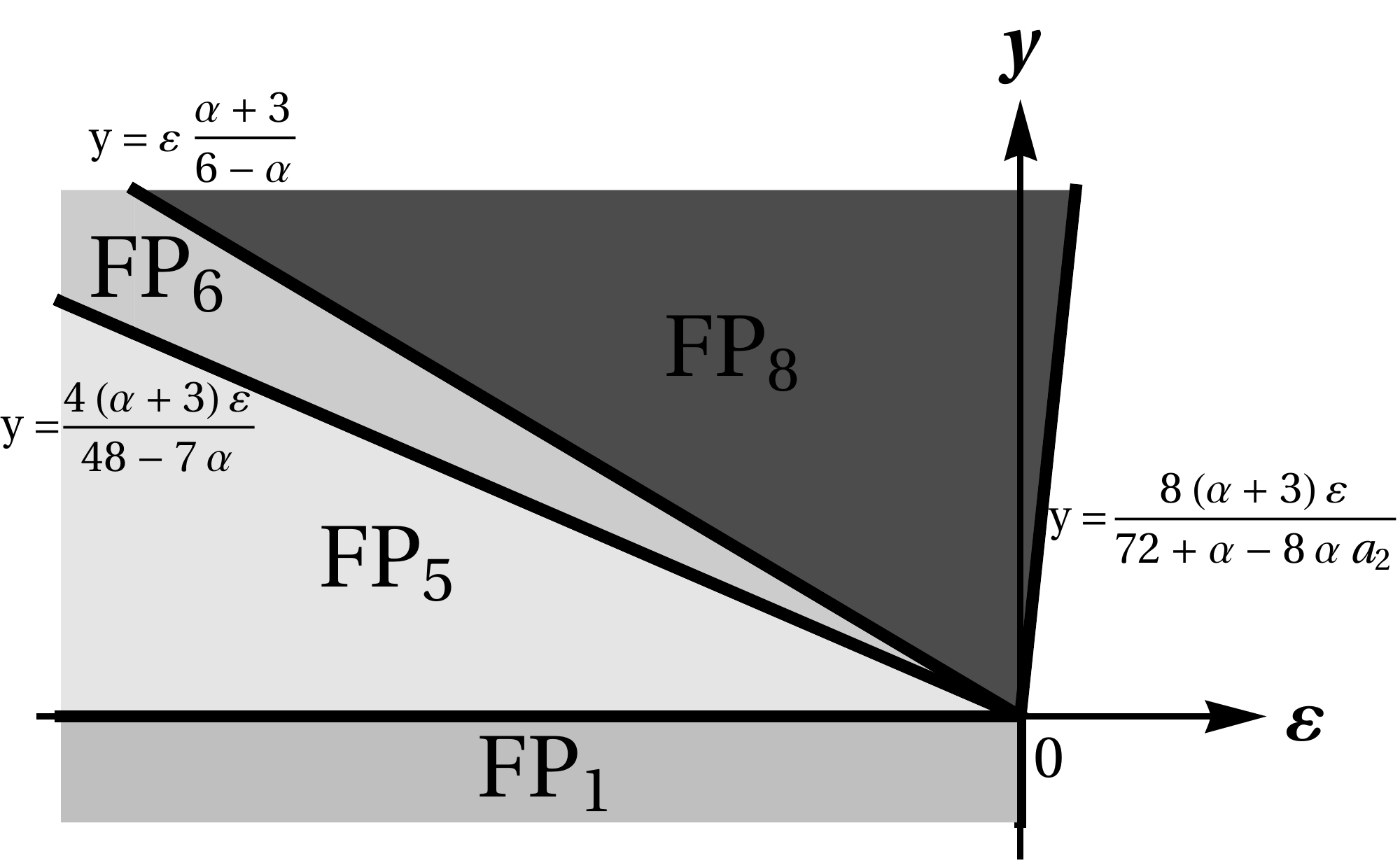}\label{fig_g}}
  \hfill
  \subfloat{\includegraphics[width=0.4\textwidth]{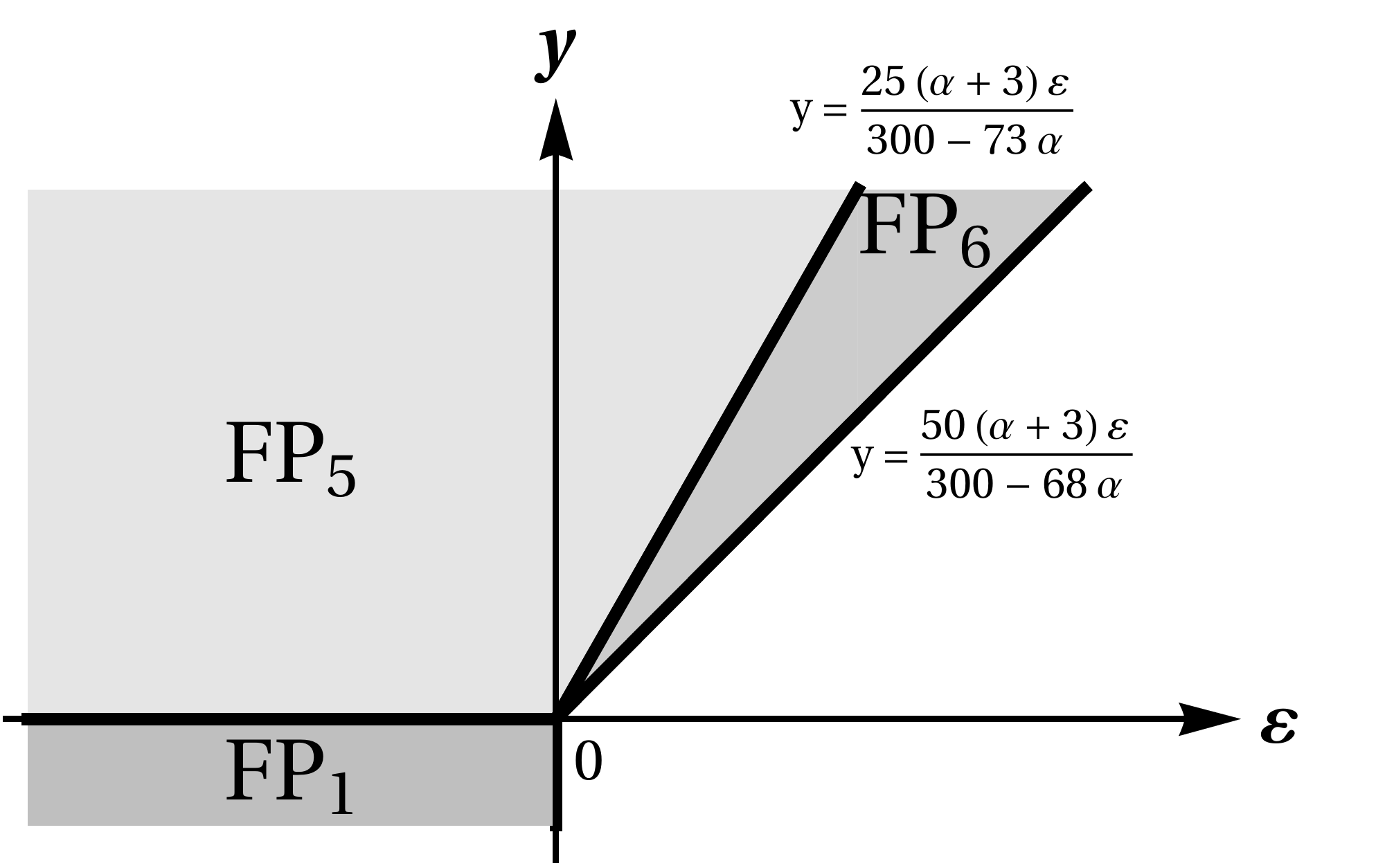}\label{fig_h}}
  \caption{Stability regions of the fixed points in the model
   for the following restricted choice of parameters
   $a_1= 1/4 $, $\alpha>8$ and $ a_2 \in (0, \frac{1}{8} + \frac{9}{\alpha})$
  (top)  
  and  
  $a_1 = 1 /10 $, $\alpha \in (\frac{50}{13},\frac{300}{73}), a_2 \in (0,\frac{9}{4} + \frac{15}{2\alpha})$
  (bottom).
  }
  \label{fig:graph4}
\end{figure}

The remaining three regimes FP6, FP7, and FP8 are possible candidates for new regimes, since for them  both velocity and
self-interactions
of model E are IR relevant. Fixed points  FP5 and FP6 differ only in IR-relevance of the self-interaction term 
$m'(\psi^{\dagger} \partial^2 \psi- \psi \partial^2 \psi^{\dagger})$, which is irrelevant for the former  and 
relevant for the latter. Similarly to the previous case of FP5, the stability region of FP6 is affected by values of free parameters
$a_1$, $a_2$ and 
$\alpha$. However, FP6 is realizable 
 only for certain intervals, which can be summarized as follows
\begin{align}
  a_1 & \in \left(0, \frac{\sqrt{3}-1}{2} \right), 
  \nonumber
  \\
  \alpha & \in \left(\frac{3}{1-2a_1-2 a_1^2}, \infty \right), 
  \\
  a_2 & \in \left(0, \frac{3(\alpha-1)}{4a_1 \alpha} + \frac{3(a_1-1)}{2} \right).
  \nonumber
\end{align}
The necessary condition for FP6 being stable is $y>0$. Further, it is restricted by the inverse inequality~(\ref{eq:FP5_a}) as
can be easily seen in Tab.~\ref{tab:eigenvalues1}. The second restriction depends on the choice of the parameter $a_2$. If 
the value of $a_2$ 
is smaller than the value 
 $a_{2b} = (3+\alpha)/(4 a_1 \alpha) + (5 a_1-1)/2$,
the second boundary is given by the inequality (\ref{eq:FP5_b}).
 For $a_2>a_{2b}$, the boundary is given by the following inequality
\begin{equation}
  [9-\alpha(1-6 a_1+6 a_1^2 +4 a_1 a_2)] y > \varepsilon (3+\alpha).
  \label{eq:fp6_restriction}
\end{equation}
 The overall
analysis of FP6 can be divided into two cases, whereby each of them corresponds to a different interval of
parameters. The situation is qualitatively
the same for 
 $a_1 > {1}/{5}$ as in previous case of FP5. For 
 $a_1 < {1}/{5}$ the region of stability 
lies in the first quadrant of $(\varepsilon,y)$-plane (see Fig.~\ref{fig_h}). 

For illustration purposes, the regions of IR stability in the $(\varepsilon,y)$-plane are depicted in 
Fig.~\ref{fig:graph4}. The regime FP6 is stable for the choice 
 $a_1= 1/4$, $\alpha>8$ and the 
border is specified by (\ref{eq:FP5_b}) up to 
 $a_2 = 9/8 + 3/\alpha$, Fig.~(\ref{fig_g}). With an increasing value of $a_2$, the 
restriction changes to (\ref{eq:fp6_restriction}) and the
 stability region
shrinks down into a boundary line for 
 $a_2 = {15}/{8} - {3}/{\alpha}$.
Beyond the value
 $a_2 = 1 + {3}/{\alpha}$ 
 the border of 
 FP5 is specified by the inequality (\ref{eq:FP5_c}) and in the $(\varepsilon,y)$-plane a void region appears
between regimes FP5 and FP6.


For the remaining two fixed points FP7 and FP8, the fixed points' coordinate of charge $g_3$ is zero.
From the results in Appendix~\ref{app:coord} it can be readily noticed that the main difference between regimes
FP5 and FP7 lies in the IR relevance of the quartic interaction terms
 (proportional to the charge $g_1$). 
 In addition, the coordinate $a_1$ attains a definite value 
  $ 1 / 4 $ for the point FP7, whereas for FP5 it is not fixed.  
  
  In contrast to a previously studied case, the stability analysis of FP7 is less involved.
  First, one can show that free parameters $\alpha$ and $a_2$ have to belong to the following intervals
\begin{equation}
\alpha \in \left(0, 8\right), \qquad a_2 \in \left(0, \frac{3}{2} \right).
\end{equation}
Further, parameter $y$ has to be strictly positive and further restrictions read
\begin{align}
  (3+\alpha) \varepsilon &> (6-\alpha) y, 
  \label{eq:FP7_a} \\
  (48-7\alpha)y &> 4(3+\alpha) \varepsilon, 
  \label{eq:FP7_b} \\
  2 [3 - \alpha(a_2-1)]y &> (3+\alpha) \varepsilon,	
  \label{eq:FP7_c}
\end{align}
 For small values of the parameter $a_2$ the second 
 restriction comes from (\ref{eq:FP7_b}) and it holds up to 
 $a_2 = 15 / 8 - 3/ \alpha$. Above this value, the
boundary is defined by inequality (\ref{eq:FP7_c}). 

Let us note that FP7 is unstable for incompressible case (Fig.~\ref{fig_a}), whereas even small (non-zero) values
 of $\alpha$ leads to
its stabilization. With
 an increasing value of $\alpha$, the region of IR stability
rotates   counterclockwise (see Fig.~\ref{fig:graph2}) and finally shrinks down to a boundary line 
 $y = - 11\eps/2$
 for the limit value of compressibility parameter $\alpha = 8$ (see Fig.~\ref{fig:graph3}).

For the remaining fixed point FP8 parameters $\alpha$ and $a_2$ are restricted by the following
conditions
\begin{equation}
  \alpha \in \left( \frac{8}{5}, \infty \right), \quad a_2 \in \left( 0, \min \bigg{\{ } 
  \frac{9}{8} + \frac{3}{\alpha}, \frac{15}{8} -\frac{3}{\alpha} \bigg{ \} }  \right).
\end{equation}
 The stability region is always bounded by the inequality 
\begin{equation}
[72-\alpha(8a_2-1)]y > 8 (3+\alpha) \varepsilon.
\end{equation}
The second restriction is given by (\ref{eq:FP7_b}) with the opposite sign of inequality for $\alpha<8$ and the 
regime is stable for 
 $a_2 < 15 / 8  - 3/\alpha$. On the other hand, for $\alpha>8$, the inequality (\ref{eq:FP7_a}) determines the
 boundary and FP8 is 
stable for  
 $a_2 < 9 / 8  + 3 / \alpha $. With an increasing value of 
parameter $\alpha$, the region of IR stability rotates counterclockwise. In the potential limit $\alpha \rightarrow \infty$ 
there always exists stability region whenever
 $a_2 < 9 / 8 $ is fulfilled. Once $a_2 >  9/8$ the regime becomes unstable.

It is worth mentioning that we have not found a non-trivial fixed point that would correspond to a case with 
all the nonlinearities being IR relevant. Notwithstanding this observation, activated
 velocity fluctuations affect the stability analysis through newly introduced charges $g_5, w, a_1$, and $a_2$. 
From the practical point of view they 
contribute to the $\Omega$ matrix (\ref{eq:omega}) with new columns and rows present. 
 Indeed, a comprehensive numerical analysis has revealed that they play an essential role
in the fixed points’ stability.

The same result has been obtained in the case of incompressible fluid \cite{PRE16}. We conclude that the presence 
of compressibility has a stabilizing effect on
the regimes where nonlinearities are relevant. The regions of IR stability for these fixed points are shown in 
Figs.~\ref{fig:graph1}-\ref{fig:graph4}.

\begin{figure}[!tbp]
  \centering 
  \subfloat{\includegraphics[width=0.4\textwidth]{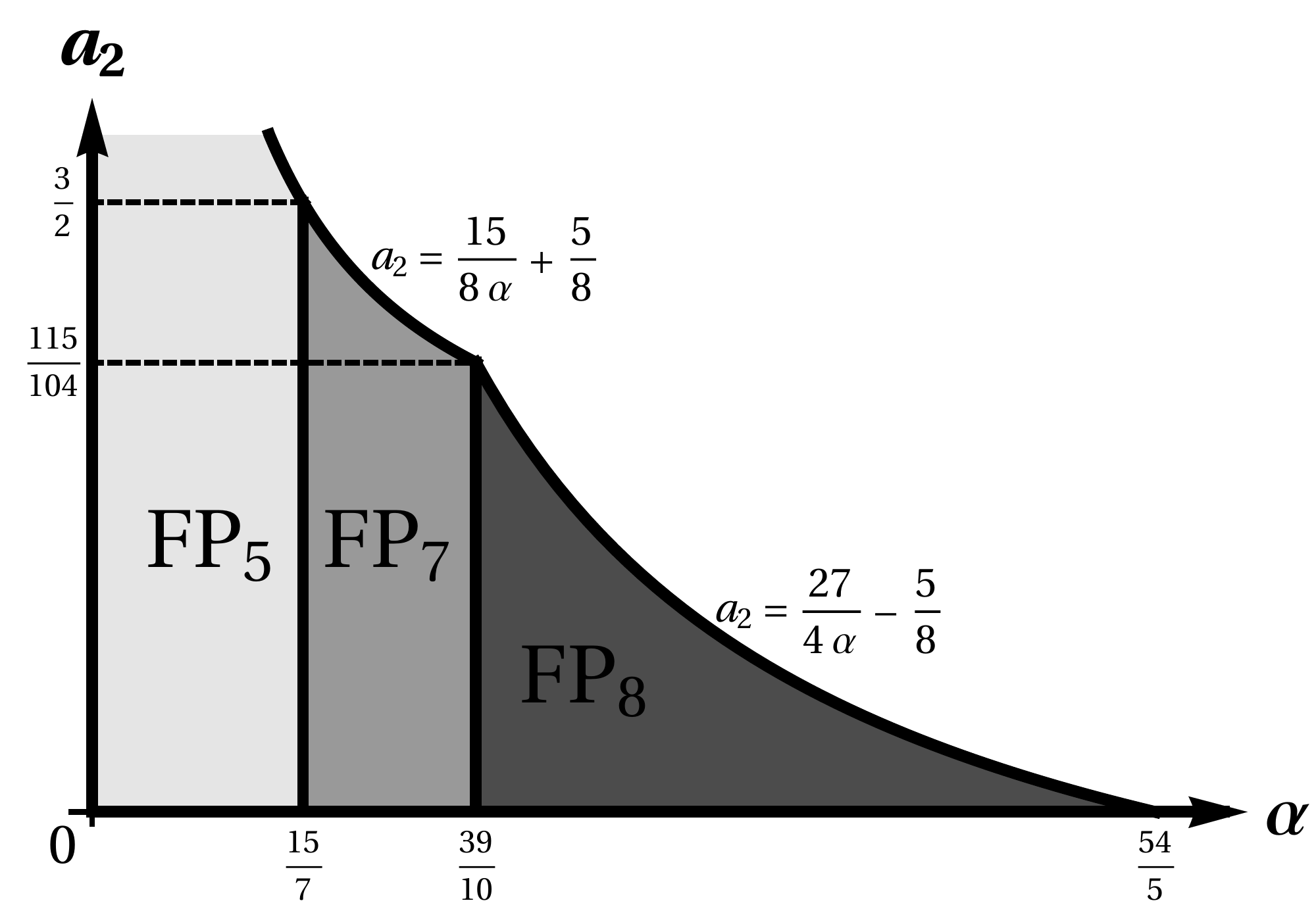}\label{fig_j}}
  \hfill
  \subfloat{\includegraphics[width=0.4\textwidth]{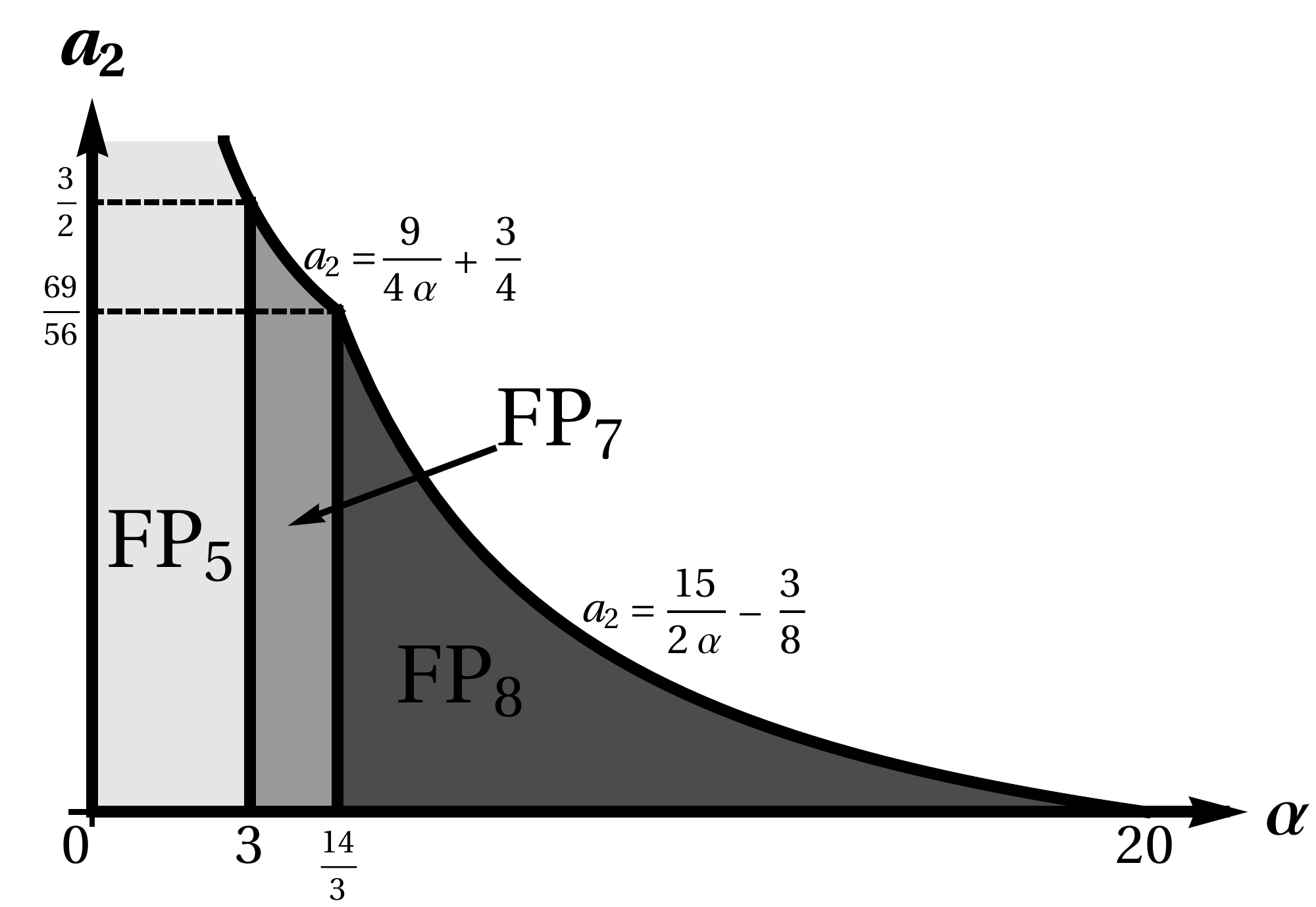}
  \label{fig_i}}
  \caption{
  Stability regions in the plane ($\alpha, a_2$) in which depicted areas correspond to the fixed points for 
  fixed parameters $\eps = 1$ and 
   $a_1 = 1 / 4 $.
   Choice $y = {4}/{3}$ corresponds to   
   the Kolmogorov spectrum of velocity
   (top).
   Choice $y=2$ corresponds to the Batchelor limit 
   (bottom).
   }
  \label{fig:graph5}
\end{figure}

Let us focus on two special cases that correspond to the Kolmogorov spectrum of the velocity $ y = 4/3$ and Batchelor 
limit $ y = 2$ (smooth velocity field), respectively.
We can see that three fixed points belong to a given value of scaling parameter $ y $ for the real space 
dimension $d = 3$ ($\varepsilon = 1$). The regime FP6 is
located in a non-physical region and could not be realized. Further, the analysis is focused on the case 
 $a_1 = 1/4$, where
the rest of the nontrivial regime is 
depicted in~Fig.~\ref{fig:graph5}.
For small values of the compressibility parameter $\alpha$ both the Kolmogorov regime  and Bachelor limit belong to universality
class FP5. As has been already mentioned 
this regime corresponds to a passively advected scalar without self-interaction and for a small value of $a_1, a_2, \alpha$, it
still can be stable for real scaling 
parameters $y, \eps$. However, for a larger value of $\alpha$, the Kolmogorov and Batchelor  value happen to lie in the 
stability region (Fig.~\ref{fig:graph5})
of the novel non-trivial regime FP7 or FP8. 

Nevertheless, we expect that a qualitative picture for large values of compressibility should remain the same.  In order
to properly describe effects of strong 
compressibility and to better understand non-universal effects for turbulent mixing one should proceed one step further 
and employ a more sophisticated model 
for compressible velocity  fluctuations \cite{VN96, ANU97,AGKL17}.

{\section{Critical dimensions} \label{sec:crit_dim}}

Existence of an IR attractive fixed point implies existence of scaling behavior of the Green functions in the 
IR range. In this critical scaling all the IR 
irrelevant parameters ($\lambda$, $\mu$ and the coupling constants) are fixed and the IR relevant parameters 
(coordinates / momenta, times~/~frequencies and 
the fields) are dilated. In the leading IR asymptotic behavior of renormalized Green 
functions $G^R$  satisfy the RG 
equation~(\ref{eq:diff_oper}) with the substitution $g\to g_{*}$ for the full set of the couplings~\cite{Vasiliev,Amit}. 
This directly yields the fundamental RG equation
\begin{equation}
  \left\{
  {\cal D}_{\mu} - \gamma_{\lambda}^{*}{\cal D}_{\lambda}
  + \sum_{\varphi} N_{\varphi}\gamma_{\varphi}^{*}  \right\} \, G^{R} = 0,
  \label{eq:RGF}
\end{equation}
where for convenience we write ${\cal D}_x \equiv x \partial/\partial x$,  $\gamma_{F}^{*}$ is the value of the
anomalous dimension at the fixed point, and
the summation over all types of the fields $\varphi$ appearing in $G^R$  is implied. Equations of this type
describe the scaling with dilatation of the
variables whose derivatives enter the differential operator \cite{Amit,Vasiliev}. 

The canonical scale invariance with respect to momentum and frequency variable, respectively, can be expressed by two relations
\begin{equation}
  \left[\sum _{\sigma}d_{\sigma}^k{\cal D}_{\sigma}-
  d_{G}^k\right]G^{R}=0 ,\quad
  \left[\sum _{\sigma}d_{\sigma}^{\omega }{\cal D}_{\sigma}-
  d_{G}^{\omega }\right]G^{R}=0 ,
  \label{eq:Canonic-Scl-Inv}
\end{equation}
where $\sigma$ is the full set of all the arguments of $G^{R}$, $d_\sigma^{k}$ and $d_\sigma^{\omega}$ are 
canonical dimensions of variable $\sigma$ with respect to momentum, and frequency, respectively. In
  order to derive proper scaling relation with fixed IR irrelevant parameters $\mu$ and $\nu$ one has to 
combine Eqs.~(\ref{eq:RGF}) and~(\ref{eq:Canonic-Scl-Inv})
in such a way that the derivatives with respect to these parameters are eliminated~\cite{turbo,Vasiliev}. This 
yields an equation of critical IR scaling for the model
\begin{equation}
  \left\{ - {\cal D}_{\bm x} + \Delta_{t} {\cal D}_{t}
  + \Delta_{\lambda} {\cal D}_{\lambda} -
  \sum_{\varphi} N_{\varphi} \Delta_{\varphi} \right\} \, G^{R} = 0  
  \label{eq:KS}
\end{equation}
with {the following notation}
\begin{equation}
  \Delta_{F} = d^{k}_{F}+ \Delta_{\omega}d^{\omega}_{F} + \gamma_{F}^{*},
  \quad \Delta_{\omega} = - \Delta_{t} =  2-\gamma_{\lambda}^{*}.
  \label{Krit}
\end{equation}
Here, $\Delta_{F}$ is the critical dimension of the quantity $F$, while $\Delta_{t}$ and $\Delta_{\omega}$ are
the critical dimensions of time and 
 frequency, and $\gamma_F^{*}$ is the value of the anomalous dimension of a quantity $F$ at the fixed point. In
our case we have obtained critical
dimensions for parameters and fields of IR stable fixed points in the following form 
\begin{itemize}
\item FP1 (Gaussian fixed point)
\begin{align}
  \Delta_{\omega} & = 2, \quad \Delta_{\psi} = \frac{d}{2}-1, \quad \Delta_{\psi'} = \frac{d}{2}+1, \\
  \Delta_m & =\Delta_{m'} = \frac{d}{2}. \nonumber
\end{align}

\item FP5 (passively advected scalar without self-interaction)
\begin{align}
  \Delta_{\omega} & = 2-y, \quad \Delta_{\psi} = 1-\frac{\varepsilon}{2}+\frac{ y }{2}-
  \frac{2\alpha y(a_1-1)^2 }{3+\alpha}, \nonumber \\ 
  \Delta_{\psi'} & = 3-\frac{\varepsilon}{2}-\frac{ y }{2}+\frac{2\alpha y (a_1-1)^2}{3+\alpha}, \\ 
  \Delta_m & = \Delta_{m'} = 2 - \frac{\varepsilon}{2}. 
  \nonumber
\end{align}

\item FP6 (new non-trivial fixed point)
\begin{align}
  \Delta_{\omega} &  = 2 - y, \quad \Delta_{\psi} = 1-\frac{\varepsilon}{2}+\frac{ y }{2}-
  2 \alpha y\frac{(a_1-1)^2}{3+\alpha}, 
  \nonumber
  \\ 
  \Delta_{\psi'} &  = 3-\frac{\varepsilon}{2}-\frac{ y }{2}+ 2\alpha y\frac{(a_1-1)^2}{3+\alpha}, \\
  \Delta_m  & = 2-\varepsilon+2y -\alpha y \frac{5+3(2a_1-1)^2}{2(3+\alpha)},   
  \nonumber \\
  \Delta_{m'} & =2-2y +\alpha y \frac{5+3(2a_1-1)^2}{2(3+\alpha)}.  
  \nonumber
\end{align}

\item FP7 (new non-trivial fixed point)
\begin{align}
  \Delta_{\omega} & = 2- y, \quad \Delta_{\psi} = 1-\frac{\varepsilon}{2} + \frac{12 - 5\alpha}{8(3+\alpha)} y, 
  \nonumber \\  
  \Delta_{\psi'} & = 3-\frac{\varepsilon}{2}-\frac{12-5\alpha}{8(3+\alpha)} y,   
  \\
  \Delta_m  & = \Delta_{m'} = 2-\frac{\varepsilon}{2}.
  \nonumber
\end{align}

\item FP8 (new non-trivial fixed point)
\begin{align}
  \Delta_{\omega} &  = 2-y, \quad \Delta_{\psi} = 1-\frac{\varepsilon}{2} + \frac{12 - 5\alpha}{8(3+\alpha)} y, 
  \nonumber
  \\ 
  \Delta_{\psi'} & = 3-\frac{\varepsilon}{2}-\frac{12-5\alpha}{8(3+\alpha)} y, 
  \\ 
  \Delta_{m'} & = 2 - \frac{48-7\alpha}{8(3+\alpha)}y, \quad
  \Delta_m  = 2-\varepsilon + \frac{48-7\alpha}{8(3+\alpha)}y. 
  \nonumber
\end{align}

\end{itemize}
{\section{Conclusion} \label{sec:conclusion} }
We have incorporated effects of compressible turbulent mixing and stirring
 in model E of critical dynamics. It has been shown how the field-theoretic formulation of such model
 can be constructed.
 A multiplicative renormalizability of the ensuing model has been proven, which 
 permits us to employ a field-theoretic perturbative renormalization group.
 Altogether $62$ nontrivial Feynman diagrams
 have been identified to the leading one-loop approximation.
 We have 
found that depending on
the values of a spatial dimension ($d=4-\varepsilon$), a scaling exponent $y$ describing statistics of 
velocity  fluctuations
and a degree of compressibility $\alpha$, the model exhibits $5$ possible large-scale regimes 
 corresponding to distinct universality classes. Two of them 
are already well-known:
 a Gaussian or trivial fixed point and a passively advected scalar without any self-interaction. 
 The remaining three regimes correspond 
 to novel universality classes for which non-linearities of model E and turbulent mixing are both relevant. 
 Critical exponents have been calculated and they exhibit dependence on $d, y$ and the 
compressibility parameter $\alpha$.
 We have found  that compressibility enhances the role of the nonlinear terms in the dynamical equations. The
stability region in the 
$(\varepsilon,y)$ plane, where new nontrivial regimes are stable is thus getting much wider as the degree of
compressibility increases. 
 As a result, turbulent mixing becomes more efficient due to combined effects of the mixing and the 
nonlinear terms.

\begin{acknowledgments}
The authors thank V. \v{S}kult\'ety for checking part of
the calculations in Appendix A and B. The authors are indebted to Nikolay M. Gulitskiy
for critical reading of the manuscript.
The work was supported by VEGA grant No. 1/0345/17 of the Ministry of Education, Science, 
Research and Sport of the Slovak Republic,
 the grant of the Slovak Research and Development Agency under the contract No. APVV-16-0186.
 \end{acknowledgments}

\appendix

{\section{Computation of Feynman diagrams}  \label{sec:app_feynman}}
In order to simplify notation we have used the following 
shift

 of coupling constants
\begin{equation}
  \frac{e S_d}{(2\pi)^2} \rightarrow e, 
\end{equation}
where $S_d =2 \pi^{\frac{d}{2}} / \Gamma({d}/{2})$ is a convenient 
geometrical factor, and
 $e\in\{g_1,g_3^2,g_3 g_5, g_5^2,w\}$. In order to avoid 
 any potential
  ambiguities
we explicitly 
indicate
 a symmetry coefficient of a given Feynman graph in front of its graphical representation. 
  The 
assessment and direction of
 external momenta for 1PI diagrams $m'\psi \psi^\dagger$
  correspond to independent momenta $\mpp$ and $\mq$ displayed
in Fig.~\ref{fig:vertex}. 
 There in vertex $ \psi^\dagger \psi m'$, external
 momenta $ \mpp,\mq $ flow in 
through fields
  $ \psi^{\dagger}(\mpp) $ and $ \psi(\mq) $
  , and flow out through  $ m'(-\mpp-\mq) $.
   The external momenta are chosen
 in a way, that they flow only via 
one internal line at most.
Further, for 1PI diagrams ${\psi^{\dagger}}'\psi \mv$ external momenta $ \mpp,\mq $ flow in as $ \psi(\mpp) $ 
and $ v_{i}(\mq) $ and then flow out as $ {\psi^{+}}'(-\mpp-\mq) $, and last
for diagrams $m'm\mv$ external momenta $ \mpp,\mq $ flow in as $ m(\mpp) $ and $ v_{i}(\mq) $ and then flow
out as $ m'(-\mpp-\mq) $. 

Let us also note that we give results only for diagrams that yield nonzero contributions.
\begin{align}
   \raisebox{-0.05cm}{\includegraphics[width=2cm]{pxsps1.pdf}}  & =\frac{2\lambda g_{3}^{2}}{(1+u)\eps}, \\
  \raisebox{-0.05cm}{\includegraphics[width=2cm]{pxsps2.pdf}} & =  \frac{\lambda (a_{1}-1)^{2}\alpha  w }{y},  \\
  \frac{1}{2} \, \raisebox{-0.05cm}{\includegraphics[width=2cm]{msms1.pdf}} & =  
   \frac{2 \lambda   g_{5}^{2} \mpp^2 }{d \eps }, \\
  \frac{1}{2} \, \raisebox{-0.05cm}{\includegraphics[width=2cm]{msms2.pdf}}& = 
    \frac{ (d-1+\alpha)\lambda w \mpp^2 }{2 d  y}, \\
  \raisebox{-0.05cm}{\includegraphics[width=2cm]{pxsp1.pdf}} & = \!\! -\!\!\left[ \frac{i\Omega}{(1+u)^{2}}\! +\! \lambda \mpp^{2} \frac{4\!-\!d(u+1) }{d(1+u)^{3}}\right]\!\!\frac{g_3^2}{\eps}, \\
  \raisebox{-0.05cm}{\includegraphics[width=2cm]{pxsp2.pdf}} & =  \frac{g_3 g_5}{\eps}\biggl[ \frac{i \Omega }{(1+u)^{2}} \nonumber\\
  & - \lambda \mpp^{2} \frac{ ( d(1+u)(2+u) - 4 ) }{d (u+1)^3 }\biggl], \\
  \raisebox{-0.05cm}{\includegraphics[width=2cm]{pxsp3.pdf}} & = - \frac{ \lambda (d-1+\alpha) w \mpp^{2} }{2d y}, \\
   \raisebox{-0.05cm}{\includegraphics[width=2cm]{msm1.pdf}} & = 
     -  \frac{(d-2) \lambda  g_{3} g_{5} \mpp^{2}}{2d  \eps}, \\
  \raisebox{-0.05cm}{\includegraphics[width=2cm]{msm2.pdf}} & = 
  -  \frac{(d-2) \lambda  g_{3} g_{5} \mpp^{2}}{2d \eps} , \\
  \raisebox{-0.05cm}{\includegraphics[width=2cm]{msm3.pdf}} & = 
   \frac{(d-1+\alpha)\lambda w \mpp^{2}}{2d  y} ,  \\ 
  \raisebox{-0.45cm}{\includegraphics[width=2.cm]{pxspm1.pdf}}  & = - \frac{i\lambda g_1 g_3}{3\eps}, \\
  \raisebox{-0.45cm}{\includegraphics[width=2.cm]{pxspm2.pdf}}  & =  \frac{i\lambda g_{1} g_{3}}{3\eps}, \\
  \raisebox{-0.8cm}{\includegraphics[width=2.cm]{pxspm3.pdf}}  & =   \frac{i\lambda g_{3}^{3}}{(1+u)^{2}\eps},  \\
  \raisebox{-0.8cm}{\includegraphics[width=2.cm]{pxspm4.pdf}}  & =   \frac{-i\lambda(3+u)g_{3}^2 g_{5}}{2(1+u)^{2}\eps},  \\
  \raisebox{-0.8cm}{\includegraphics[width=2.cm]{pxspm5.pdf}}  & =   \frac{i\lambda g_{3}^2 g_{5}}{2(1+u)\eps}, \\
  \raisebox{-0.8cm}{\includegraphics[width=2.cm]{pxspm6.pdf}}  & = -  \frac{i\lambda g_{3} \alpha  a_{1} a_{2} w}{(1+u)y}, \\
  \raisebox{-0.45cm}{\includegraphics[width=2.cm]{pxpms1.pdf}}  & = \frac{i\lambda g_{5}g_1}{3d\eps} \biggl[ {(6-d)\mpp^{2}} + {(d-2)\mq^{2}} \nonumber\\
  & + {4\mpp\cdot\mq}  \biggl],   \\
  \raisebox{-0.45cm}{\includegraphics[width=2.cm]{pxpms2.pdf}}  & = -\frac{i\lambda g_{5} g_1}{3d\eps} \biggl[
     {(6-d)\mpp^{2}} + {(d-2)\mq^{2}} \nonumber\\
     & + {4\mpp\cdot\mq}  \biggl], \\
  \raisebox{-0.9cm}{\includegraphics[width=2.cm]{pxpms3.pdf}}  & = i\lambda g_{5} (\mpp^{2}\! - \!\mq^{2}) \frac{[d-4+(d-2)u] g_{3}^{2}}{d(1+u)^{2} \eps},  \\
  \raisebox{-0.9cm}{\includegraphics[width=2.cm]{pxpms4.pdf}}  & = \frac{i\lambda g_{5}^2 g_3}{\eps} \biggl[  \frac{(6-d+(2-d)u)\mq^{2}}{2d(1+u)^{2}} 
  \nonumber \\
  & +
  \frac{(d-2)\mpp^{2}}{2d(1+u)} + 
  \frac{4\mpp\cdot\mq}{2d(1+u)^{2}} \biggl], \\
  \raisebox{-0.9cm}{\includegraphics[width=2.cm]{pxpms5.pdf}}  & = -\frac{i\lambda g_{5}^2 g_3}{\eps} \biggl[
  \frac{(d-2)\mq^{2}}{2d(1+u)} + \frac{4\mpp\cdot\mq}{2d(1+u)^{2}} 
  \nonumber \\
  & + \frac{(6-d+(2-d)u)\mpp^{2}}{2d(1+u)^{2}} \biggl], \\
  \raisebox{-0.9cm}{\includegraphics[width=2.cm]{pxpms6.pdf}}  & =(\mpp^{2}-\mq^{2}) \frac{ i\lambda g_{5} \alpha a_{1} (2+(d-2)a_{1}) w}{2dy},  \\
  \frac{1}{2}\, \raisebox{-0.45cm}{\includegraphics[width=2.cm]{pxspxpp1.pdf}} & = 
  \frac{\lambda g^2_{1}}{9\eps}, \\
  \raisebox{-0.45cm}{\includegraphics[width=2.cm]{pxspxpp2.pdf}} & = \frac{2\lambda g_{1}^2}{9\eps }, \\
  \raisebox{-0.45cm}{\includegraphics[width=2.cm]{pxspxpp3.pdf}} & =  \frac{2\lambda g_{1}^2}{9\eps}, \\
  \frac{1}{2}\,\raisebox{-0.9cm}{\includegraphics[width=2.cm]{pxspxpp4.pdf}} & = \frac{ \lambda g_1 g_{3}^{2}}{3(1+u)\eps}, \\
  \raisebox{-0.9cm}{\includegraphics[width=2.cm]{pxspxpp5.pdf}} & = - \frac{2\lambda g_1 g_{3}^{2}}{3(1+u)\eps}, \\
  \raisebox{-0.9cm}{\includegraphics[width=2.cm]{pxspxpp6.pdf}} & =  \frac{2\lambda g_1 g_{3}^{2}}{3(1+u)^{2} \eps} , \\
  \frac{1}{2}\, \raisebox{-0.9cm}{\includegraphics[width=2.cm]{pxspxpp7.pdf}} & =  -\frac{\lambda g_1 g_{3}^{2}}{3(1+u)^{2}\eps}, \\
  \raisebox{-0.9cm}{\includegraphics[width=2.cm]{pxspxpp8.pdf}}& =   \frac{\lambda g_1 g_{3}g_{5}}{3(1+u)\eps} , \\
  \raisebox{-1.1cm}{\includegraphics[width=2.cm]{pxspxpp9.pdf}} & =   -\frac{\lambda g_1 g_{3} g_{5}}{3(1+u)\eps}, \\
  \raisebox{-0.9cm}{\includegraphics[width=2.cm]{pxspxpp10.pdf}} & =   \frac{\lambda g_1g_{3}g_{5}}{3(1+u)\eps}, \\
  \frac{1}{2}\, \raisebox{-0.9cm}{\includegraphics[width=2.cm]{pxspxpp11.pdf}} & = \frac{\lambda (3+u) g_1 g_{3}g_{5}}{6(1+u)^{2}\eps},  \\
  \raisebox{-0.9cm}{\includegraphics[width=2.cm]{pxspxpp12.pdf}} & =  -\frac{\lambda(3+u)g_1 g_{3}g_{5}}{3(1+u)^{2} \eps}, \\
  \raisebox{-0.9cm}{\includegraphics[width=2.cm]{pxspxpp13.pdf}} & = 
  -\frac{\lambda g_1 g_{3}g_{5}}{ 3(1+u)\eps}, \\
  \frac{1}{2}\, \raisebox{-0.9cm}{\includegraphics[width=2.cm]{pxspxpp14.pdf}} & =  
  \frac{\lambda g_1 g_{3}g_{5}}{ 6(1+u)\eps} , \\
  \raisebox{-0.9cm}{\includegraphics[width=2.cm]{pxspxpp15.pdf}} & =  \frac{\lambda(1+2u) g_{3}^{3}g_{5}}{2u(1+u)^{2} \eps} , \\
  \raisebox{-0.9cm}{\includegraphics[width=2.cm]{pxspxpp16.pdf}} & = \frac{\lambda (1+2u)g_{3}^{3}g_{5}}{2u(1+u)^{2}\eps}, \\
  \raisebox{-0.9cm}{\includegraphics[width=2.cm]{pxspxpp17.pdf}} & =  -\frac{\lambda(2+u) g_{3}^{2}g_{5}^{2}}{2u(1+u)^{2} \eps}, \\
  \raisebox{-0.9cm}{\includegraphics[width=2.cm]{pxspxpp18.pdf}} & =   -\frac{ \lambda (2+u) g_{3}^{2}g_{5}^{2}}{2u(1+u)^{2}\eps}, \\
  \raisebox{-0.9cm}{\includegraphics[width=2.cm]{pxspxpp19.pdf}} & =  \frac{\lambda g_{3}^{3}g_{5}}{2u(1+u)^{2} \eps}, \\
  \raisebox{-0.9cm}{\includegraphics[width=2.cm]{pxspxpp20.pdf}} & = \frac{\lambda g_{3}^{3}g_{5}}{2u(1+u)^{2} \eps }, \\
  \raisebox{-0.9cm}{\includegraphics[width=2.cm]{pxspxpp21.pdf}} & =  -\frac{\lambda g_{3}^{2}g_{5}^{2}}{2(1+u)^{2} \eps}, \\
  \raisebox{-0.9cm}{\includegraphics[width=2.cm]{pxspxpp22.pdf}} & =  - \frac{ \lambda g_{3}^{2}g_{5}^{2}}{2(1+u)^{2} \eps}, \\
  \frac{1}{2}\, \raisebox{-0.9cm}{\includegraphics[width=2.cm]{pxspxpp23.pdf}} & = 
  - \frac{\lambda g_1 \alpha a_{1}^{2} w}{6y}, \\
  \raisebox{-0.9cm}{\includegraphics[width=2.cm]{pxspxpp24.pdf}} & =  - \frac{\lambda g_1 \alpha a_{1}^{2} w}{3y},\\
  \raisebox{-0.9cm}{\includegraphics[width=2.cm]{pxspxpp27.pdf}} & = - \frac{\lambda \alpha a_{1}^{2}w g_{3}g_{5}}{2(1+u) y}, \\
  \raisebox{-0.9cm}{\includegraphics[width=2.cm]{pxspxpp28.pdf}} & =  \frac{\lambda \alpha a_{1}^{2}w g_{3}g_{5}}{2(1+u) y}, \\
  \raisebox{-0.57cm}{\includegraphics[width=2.5cm]{pxspv2.pdf}} & = -i  q_{1} \frac{1-d a_{1}}{3d  \eps} g_{1},\\
  \raisebox{-0.57cm}{\includegraphics[width=2.5cm]{pxspv2.pdf}} & = -i  q_{1}  \frac{1-d a_{1}}{3d  \eps} g_{1},\\
  \raisebox{-1cm}{\includegraphics[width=2.5cm]{pxspv3.pdf}} & = ip_{1}  \frac{4u g_{3}^{2}}{d(1+u)^{3}\eps}  
  \nonumber \\ 
  & - i  q_{1} \frac{[2-d(1+u)a_{1}] g_{3}^{2}}{d(1+u)^{3}  \eps} ,\\
  \raisebox{-1cm}{\includegraphics[width=2.5cm]{pxspv4.pdf}} & = -ip_{1}  \frac{u(5+u) g_{3}g_{5}}{d(1+u)^{3}\eps} 
  \nonumber \\ 
  & - \frac{i  q_{1} g_3 g_5}{2d(1+u)^{3} \eps} [a_1 d \left(u^2+4 u+3\right)
  \nonumber\\ 
  & -(u^2+4 u+7)] ,\\
  \raisebox{-1cm}{\includegraphics[width=2.5cm]{pxspv5.pdf}} & = ip_{1} \frac{ug_{3}g_{5}}{d(1+u)^{2}\eps}  
   - \frac{ i  q_{1} g_3 g_5 }{2d(1+u)^{2}  \eps}
   \nonumber \\ 
  & \times \left[5 + 3u + d (u+1)(a_{1} - 1)\right]  ,\\
  \raisebox{-1cm}{\includegraphics[width=2.5cm]{pxspv6.pdf}} & = -ip_{1} \frac{4 g_{3}g_{5}}{d(1+u)^{3}\eps} 
  \nonumber \\ 
  & - i  q_{1} \frac{[-2u+d(1+u)a_{2}] g_{3} g_{5}}{d(1+u)^{3}  \eps} ,\\
  \raisebox{-1cm}{\includegraphics[width=2.5cm]{pxspv7.pdf}} & = ip_{1}  \frac{(1+5u) g_{3}^{2}}{du(1+u)^{3} \eps} 
   - \frac{ i  q_{1} g_3^2 }{2du(1+u)^{3}  \eps}
   \nonumber \\ 
  & \times [1+4u+7u^{2}\! - \! d (1 + 4 u + 3 u^{2} ) a_{2} ] ,\\
  \raisebox{-1cm}{\includegraphics[width=2.5cm]{pxspv8.pdf}} & = 
  i  q_{1} \frac{\left[ 1 + 3u + d (u+1) (a_{2} -1 ) \right] g_{3}^{2}}{2du(1+u)^{2} \eps}
  \nonumber \\ 
  & -ip_{1} \frac{g_{3}^{2}}{du(1+u)^{2}\eps},\\
  \raisebox{-1cm}{\includegraphics[width=2.5cm]{msmv1.pdf}} & = i p_{1}  \frac{g_{3} g_{5}}{2 d\eps}  
  +i q_{1}   \frac{g_{3} g_{5}}{2 d  \eps} ,\\
  \raisebox{-1cm}{\includegraphics[width=2.5cm]{msmv2.pdf}} & = i p_{1}  \frac{g_{3} g_{5}}{2d\eps}  
  +i q_{1}   \frac{g_{3} g_{5}}{2 d  \eps} ,\\
  \raisebox{-1cm}{\includegraphics[width=2.5cm]{msmv3.pdf}} & = i p_{1}  \frac{g_{3} g_{5}}{2 d\eps}
  +i q_{1}   \frac{g_{3} g_{5}}{2 d  \eps} ,\\
  \raisebox{-1cm}{\includegraphics[width=2.5cm]{msmv4.pdf}} & = i p_{1}  \frac{g_{3} g_{5}}{2 d\eps} 
  +i q_{1}   \frac{g_{3} g_{5}}{2 d  \eps} ,\\
  \raisebox{-1cm}{\includegraphics[width=2.5cm]{msmv5.pdf}} & = -i p_{1}  \frac{g_{3} g_{5}}{d\eps} 
  -i q_{1}  \frac{g_{3} g_{5}}{d \eps} ,\\
  \raisebox{-1cm}{\includegraphics[width=2.5cm]{msmv6.pdf}} & = -i p_{1}  \frac{g_{3} g_{5}}{d\eps}
  -i q_{1}   \frac{g_{3} g_{5}}{d  \eps}.
\end{align}

{\section{Anomalous dimensions to the one-loop order } \label{sec:app_anomdim}}
In this section, we review the explicit expressions
for the anomalous dimension $\gamma_x$, $x\in\{g_1,g_3,g_5,u,w,a_1,a_2\}$ of the
charges and for the fields $x\in\{\psi,\psi^\prime,\mv\}$, respectively.
From  relations~\eqref{eq:rg_constants1}-\eqref{eq:rg_constants5}, the following expressions directly follow
\begin{align}
\gamma_{\lambda} & = -\gamma_w =  \frac{4 g_3^2}{d(1+u)^3} + \frac{g_3g_5[d - 4 + d u(2+u)]}{d (1+u)^3} \nonumber \\
& + \frac{w(d - 1+\alpha)}{2 d}, \\
\gamma_u & = - \frac{4 g_3^2}{d(1+u)^3} + \frac{w(d - 1 + \alpha)(1-u)}{2du}  \nonumber \\ 
& - \frac{g_3g_5[ 2u^3 - u^2 (d-6) - 2u (d-1) + 2-d]}{d u(1+u)^3}, \\
\gamma_{g_3} & = - \frac{4 g_3^2}{d(1+u)^3} + \frac{g_5^2}{d u} - \frac{w}{2}\left( 1- \frac{1-\alpha}{d} - \frac{2a_1a_2 \alpha}{1+u} \right) \nonumber \\ & + \frac{g_3g_5[2(u^3+3u^2 +7 u +1)-d(1+u)^2(1+3u)]}{2du(1+u)^3}, \\
\gamma_{g_5} & = \frac{2 g_3^2[d(2+3u+u^2)-6-3u-u^2]}{d(1+u)^3}  \nonumber \\ 
& + \frac{w}{d}\left[1-d + \alpha \left(\frac{d-2}{2} + a_1 (a_1-1)(d-1)\right) \right]\nonumber\\
& + \frac{g_3g_5}{2 d u(1+u)^3}
\biggl[2(u^3+3u^2+9u-1)\nonumber\\
&-d(5u^3+13u^2+7u-1)\biggl] - \frac{g_5^2}{du},  \\
\gamma_{g_1} & = \frac{2g_3[d(2+3u+u^2)- 4](g_3 \! - \! g_5)}{d (1+u)^3} - \frac{6g_3^2g_5(g_3\! -\! g_5)}{ug_1(1+u)} 	
\nonumber \\
&  \!-\frac{5g_1}{3}  + \frac{w}{d}\left[ 1 - d + \alpha\left(\frac{d}{2} - 1 + d a_1(2a_1 -1)\right)\right],  	\\
\gamma_m & = -\gamma_{m^\prime} = \frac{g_3g_5(d-2)}{2du} - \frac{g_5^2}{du}, \\
\gamma_{\psi} &  = \gamma_{\psi^\dagger} = \frac{g_3(g_5-g_3)[d(3+4u+u^2) - 4]}{2 d (1+u)^3} \nonumber\\
& + \frac{w}{4d} \left[d - 1 -\alpha \left(d(a_1-1)^2 - 1\right)\right], \\
\gamma_{\psi^\prime} & = \gamma_{ {\psi^\dagger}^\prime} = \frac{g_3(g_5-g_3)[4-d(1+u)^2]}{2d(1+u)^3} \nonumber\\
& + \frac{w}{4d} \left[1-d + \alpha \left(d(a_1-1)^2-1\right)\right], \\
\gamma_{a_1} & = \frac{g_1(1- d a_1)}{6 a_1} + g_3 g_5\frac{d(2+u)-4}{4(1+u)^2} \nonumber \\
 & + g_3^2 \frac{4 u(1+2a_1) + d[1+u -2 a_1 u - 2a_2(1+2 u)]}{8 u(1+u)^2a_1},\nonumber  \\
 & + \frac{g_3 g_5}{8(1+u)^2a_1}\biggl[ d(2 a_2 - 1 - u) \nonumber\\
 & + 2(u - 1) + 2 a_1 [d (2+u) - 4]\biggl],\\ 
 	\gamma_{a_2} & = 0.
 	\label{eq:gamma1}
\end{align}

{ \section{ Coordinates of fixed points } \label{app:coord}}
In this section we list coordinates of all fixed points for model E with compressible velocity fluctuations. The expression
	''not fixed'' (NF) stands for a situation when a given fixed point coordinate can not be unambiguously
	determined from a solution to RG flow equations (\ref{eq:zero_beta}). 
\begin{table}[h!]
	\small\addtolength{\tabcolsep}{-2pt}
	\renewcommand{\arraystretch}{1.75}
	\begin{tabular}{|c|c|c|c|c|c|c| c |}
		\hline
		FP           & FP1     & FP2         &      FP3 & FP4 & FP5 & FP6\\ \hline
		$g_1$        & $0$     & $\frac{3\eps}{5}$           & $0$ & $\frac{3}{5}\varepsilon$ & $0$  & $0$  \\ \hline
		$g_3$        & $0$     & $0$       & $\sqrt{\eps}$ & $\sqrt{\eps}$ & $0$ & $0$ \\ \hline
		${g_5}$ & $0$     & $0$         
		& $\sqrt{\eps}$  & $\sqrt{\eps}$ & $0$ & $\sqrt{2\eps+\frac{8[\alpha-3+3a_1\alpha(a_1-1)]}{\alpha+3}y}$ \\ \hline
		$w $  & $0$     & $0$  & $0$ & $0$ &
		$\frac{8y}{3+\alpha}$ & $\frac{8y}{3+\alpha}$ \\ \hline
		$u$          & NF     & NF  & $1$ & $1$ & $1$  & $1$  \\ \hline  
		$a_1$        & NF     & $\frac{1}{4}$   
		& $a_2-\frac{1}{4}$  & $5a_2-\frac{9}{4}$ & NF & NF \\ \hline
	\end{tabular}
	\caption{Coordinates of fixed points FP1 up to FP6.}
	\label{tab:fixed_points1}
\end{table}

\begin{table}[h!]
	\small\addtolength{\tabcolsep}{-1pt}
	\renewcommand{\arraystretch}{1.75}
	\begin{tabular}{|c|c|c|c|c|}
		\hline
		FP                & FP7         &      FP8  \\ \hline
		$g_1$           & $\frac{3}{5}\left[\eps + \frac{\alpha-6}{3+\alpha}y\right]$           
		& $\frac{3}{5}\left[\eps + \frac{\alpha-6}{3+\alpha}y\right]$ \\ \hline
		$g_3$             & $0$       & $0$  \\ \hline
		${g_5}$      & $0$         
		& $\sqrt{2\eps + \frac{7\alpha-48}{6+2\alpha}y } $   \\ \hline
		$w $       & $\frac{8y}{3+\alpha}$  & $\frac{8y}{3+\alpha}$ \\ \hline
		$u$             & $1$          & $1$ \\ \hline  
		$a_1$            & $\frac{1}{4}$   
		& $\frac{1}{4}$   \\ \hline
	\end{tabular}	
	\caption{Coordinates of fixed points FP7 and FP8.}
	\label{tab:fixed_points2}
\end{table}
Fixed point's value of the charge $a_2$ is in general not fixed, only for FP3 and FP4 there is
 an aforementioned relationship between $a_1^*$ and $a_2^*$.

{ \section{ Eigenvalues of $\Omega$-matrix } \label{app:eigenvalues}}
In this section we list all eigenvalues for fixed points from App.~\ref{app:coord}.

\begin{table}[h!]
	\small\addtolength{\tabcolsep}{-1pt}
	\renewcommand{\arraystretch}{1.75}
	\begin{tabular}{|c|c|c|c|c|c|}
		\hline
	      FP1     & FP2         &      FP3 & FP4 & FP5 \\ \hline
		  $-\eps$ & $\eps$ & $-\eps$ & $-\frac{\eps}{10}$ & 
		  $2y\frac{3-\alpha+3a_1\alpha(1- a_1)}{3+\alpha}-\frac{\eps}{2}$   \\ \hline        
		  $-y$ & $-\frac{\eps}{2}$ & $-\frac{\eps}{2}$ & $\frac{\eps}{4}$ & 
		  $2 y \frac{3-\alpha+4a_1\alpha(1-2a_1)}{3+\alpha}-\eps$ \\ \hline
		  $-\frac{\eps}{2}$ & $-\frac{\eps}{2}$ & $\frac{\eps}{4}$ & $\eps$ &
		  $y \frac{3+\alpha-4a_1 a_2 \alpha}{3+\alpha}-\frac{\eps}{2}$  \\ \hline
		  $-\frac{\eps}{2}$ & $\frac{2}{5}\eps$ & $\eps$ & $\eps$ & $y$  \\ \hline
		  $0$ & $-y$ & $\frac{3\eps}{2}$ & $\frac{3\eps}{2}$ & $y$  \\ \hline
		  $0$ & $0$ & $\frac{\eps}{2}-y$ & $\frac{\eps}{2}-y$ & $0$   \\ \hline
	\end{tabular}
	\caption{Eigenvalues of matrix  $\Omega$ (see Eq.~\eqref{eq:omega}) for IR stable fixed points corresponding to regimes FP1 up to FP5.}
	\label{tab:eigenvalues1}
\end{table}

\begin{table}[h!]
	\small\addtolength{\tabcolsep}{-2pt}
	\renewcommand{\arraystretch}{1.75}
	\begin{tabular}{|c|c|c|c|c|}
		\hline
		 FP6   & FP7 & FP8 \\ \hline
		    $\eps +4y \frac{\alpha-3+3a_1\alpha(a_1-1)}{3+\alpha}$ 
		  & $y$  & $y$  \\ \hline
		   $2y\frac{3-\alpha+4a_1\alpha(1-2a_1)}{3+\alpha} - \eps$ 
		  & $y$  & $y$ \\ \hline
		  $\frac{9-\alpha + 6a_1\alpha(1-a_1)-4a_1 a_2 \alpha}{3+\alpha}-\eps$
		   & 
		  $\frac{2}{5}\left(\eps + y \frac{\alpha-6}{3+\alpha}\right)$ & 
		  $\frac{2}{5}\left(\eps + y \frac{\alpha-6}{3+\alpha}\right)$ \\ \hline
		 $y$ 
		  & $\eps + y \frac{\alpha-6}{3+\alpha}$  & 
		  $\eps + y \frac{\alpha-6}{3+\alpha}$ \\ \hline
		  $y$
		   & $\frac{1}{8}\left( \frac{48-7\alpha}{3+\alpha}y-4\eps\right)$   
		   & $\frac{7\alpha-48}{4(3+\alpha)}y + \eps$    \\ \hline
	        $0$
	        & $y-\frac{a_2 \alpha y}{3+\alpha}-\frac{\eps}{2}$ 
	        & $y\frac{72+\alpha-8a_2\alpha}{8(3+\alpha)}-\eps$  \\ \hline
	\end{tabular}	
	\caption{Eigenvalues of  matrix  $\Omega$ (see Eq.~\eqref{eq:omega}) for IR stable fixed points corresponding to regimes FP6 up to FP8.}
	\label{tab:eigenvalues2}
\end{table}

\bibliographystyle{apsrev}
\bibliography{mybib}

\end{document}